\documentclass[prl,aps,twocolumn,showpacs,preprintnumbers,amsmath,amssymb]{revtex4}
\usepackage{epsfig}
\usepackage{color}
\usepackage{ifthen}
\usepackage{graphicx}
\usepackage{amsmath}
\usepackage{amssymb}
\usepackage{xfrac}

\usepackage{graphicx}
\usepackage{dcolumn}
\usepackage{bm}
\def\Dsl{\hbox{/\kern-.6700em\it D}} 
\def\dsl{\hbox{/\kern-.5300em$\partial$}}

\def\eqa{\begin{eqnarray}}
\def\eeqa{\end{eqnarray}}
\def\eq{\begin{equation}}
\def\eeq{\end{equation}}
\def\be{\begin{equation}}
\def\ee{\end{equation}}
\def\bea{\begin{eqnarray}}
\def\eea{\end{eqnarray}}

\newcommand{\dslash}{\not{\hbox{\kern-2pt $\partial$}}}
\newcommand{\pslash}{\not{\hbox{\kern-2.3pt $p$}}}
 \newtoks\nslashfraction
 \nslashfraction={.13}
 \newcommand{\nslash}[1]{\setbox0\hbox{$ #1 $}
   \setbox0\hbox to \the\nslashfraction\wd0{\hss \box0}/\box0 }

\begin{document}

\preprint{}

\title{Model-Independent Signatures of New Physics in non-Gaussianity}
\author{Mark G. Jackson$^{1}$ and Koenraad Schalm$^2$}
\affiliation{$^1$Paris Centre for Cosmological Physics and Laboratoire AstroParticule et Cosmologie, Universit\'{e} Paris 7-Denis Diderot,}
\affiliation{B\^{a}timent Condorcet, 10 rue Alice Domon et L\'{e}onie Duquet, 75205 Paris, France}
\affiliation{$^2$Instituut-Lorentz for Theoretical Physics, University of Leiden, Leiden 2333CA, The Netherlands}

\date{\today}

\pacs{04.62.+v, 98.80.-k, 98.70.Vc}

\begin{abstract}
\noindent
We compute the model-independent contributions to the primordial bispectrum and trispectrum in de Sitter space due to high-energy physics.  We do this by coupling a light inflaton to an auxiliary heavy field, and then evaluating correlation functions in the Schwinger-Keldysh ``in-in" formalism.  The high-energy physics produces corrections parametrized by $H/M$ where $H$ is the scale of inflation and $M$ is the mass of the heavy field.  The bispectrum peaks near the elongated shape, but otherwise contains no features.  The trispectrum receives no corrections at order $H/M$.
\end{abstract}

\maketitle

\section{Introduction}
The leading paradigm for understanding the first moments of the Universe is known as Inflationary Theory, wherein space rapidly expanded in a very brief period of time \cite{ Guth:1980zm, Linde:1981mu, Albrecht:1982wi, Linde:1983gd}.  An important property of inflation is that its low-energy observables are sensitive to high-energy physics \cite{Brandenberger:1999sw, Greene:2005aj}.  This non-decoupling of energy scales, while sometimes confounding, can also be used as a probe of ultra high-energy physics such as the effects of quantum gravity.  In order to verify, constrain or eliminate classes of such theories, one must understand the precise relationship between low- and high-energy physics in inflationary theory. 

The primary low-energy inflationary observables are the correlation functions of primordial perturbations of the quantum field responsible for the inflation.  Since such perturbations seed the observed temperature fluctuations in the cosmic microwave background (CMB), one could use the CMB to gain insight into high-energy physics.  This is particularly exciting given the recent explosion in precision data by \emph{WMAP} \cite{wmap7}, \emph{Planck}  \cite{planck}, \emph{Euclid} \cite{euclid} and possibly \emph{CMBPol/Inflation Probe} \cite{cmbpol}.

The inflationary fluctuation spectrum has been observed to be nearly Gaussian \cite{wmap7}, and the two-point correlation probes the free theory:
\[  \langle \varphi_{\bf k} \varphi_{\bf p} \rangle \sim  P_\varphi(k) (2 \pi)^3 \delta^3( {\bf k} + {\bf p}), \]
 Higher-point correlations then probe the interactions by measuring the {\it non-Gaussianity} of the statistics.  The three-point correlation is commonly referred to as the bispectrum and is simply the 3-field analog of the power spectrum:
\begin{equation}
\label{bdef}
\langle \varphi_{{\bf k}_1}  \varphi_{{\bf k}_2}  \varphi_{{\bf k}_3} \rangle \sim B_\varphi({\bf k}_1,{\bf k}_2,{\bf k}_3) (2 \pi)^3 \delta^3 \left( {\bf k}_1+{\bf k}_2+{\bf k}_3 \right) .
\end{equation}
This diagnostic has quickly become an essential tool in constraining inflation models \cite{Komatsu:2009kd}.  While as yet there have only been hints of non-Gaussianity \cite{Yadav:2007yy, Senatore:2009gt}, a definitive measurement will occur soon from \emph{Planck} \cite{planck}.  At the most elementary level, non-Gaussianity can be categorized according to the shape of the triangle formed by the fields' momentum vectors at which the amplitude $B_{\varphi}$ is maximized.  The `squeezed' shape, $k_1 \sim k_2$ and $k_3 \sim 0$, dominates for local interactions such as $\varphi(x)^3$ \cite{Komatsu:2001rj}, and occur well outside the horizon.  The `equilateral' shape $k_1 \sim k_2 \sim k_3$ occur primarily at the time of horizon-crossing and dominates for higher-derivative interactions such as found in the Dirac-Born-Infeld theory \cite{Creminelli:2003iq, Alishahiha:2004eh}.  The `elongated' shape, where $k_1 + k_2 \sim k_3$, dominates for an initial vacuum state modified away from Bunch-Davies, and often enhances the other types as well \cite{Holman:2007na, Meerburg:2009ys, Meerburg:2009fi,Chen:2010bka,Ganc:2011dy, Chialva:2011hc}.  A nice heuristic way of understanding this relationship is that the more squeezed the triangle, the later it originated.  In anticipation of the upcoming \emph{Planck} data, a more sophisticated technique has recently been pioneered by Fergusson, Liguori, and Shellard \cite{Fergusson:2010dm}.  This analysis is model-independent in the sense that it does not assume a particular template shape (squeezed, equilateral, or elongated) and so requires an equally sophisticated model-independent theoretical basis for understanding.

In order to meet this need, the fundamental question how to construct the low-energy effective action in inflationary backgrounds had to be answered.  We did so in \cite{Jackson:2010cw, Jackson:2011qg,Jackson:2012qp}.  This new method allows one to reliably compute universal corrections to primordial correlation functions.  The approach is model-independent in the sense that we begin with the general form of an ultraviolet-complete action and obtain an answer in terms of a small number of free parameters that in principle encodes all the possible unknown physics.  The fundamental organizing principle is not yet known.  Nevertheless we can simply compute the effects in any specific model and generalize from there.  We have previously used this technique to compute the primordial power spectrum, here we will apply it to compute the 3- and 4-point correlations.  An important difference with flat space effective field theory is that a heavy field of mass $M$ will correct the correlation function already at order $H/M$, where $H$ is the scale of inflation.  Moreover, as will shall show, the leading contribution to the bispectrum is of the elongated shape just as previous studies surmised \cite{Holman:2007na, Meerburg:2009ys, Meerburg:2009fi,Ganc:2011dy, Chialva:2011hc}.

This article is structured as follows.  In \S2 we briefly summarize the technical framework by which high-energy physics can consistently be integrated out of a theory in an inflating background.  In \S3 we calculate the bispectrum corrections, and in \S4 those for the trispectrum.  In \S5 we conclude and offer some remarks about detectability and future directions.

\section{Setup}

\subsection{The Generic Action}
We will consider the most general field theory containing a light scalar field fluctuation $\varphi$ renormalizably coupled to a heavy field $\chi$:
\begin{eqnarray}
\nonumber
S_{\rm fl} &=& \int d^4 x \sqrt{g} \left[ \frac{1}{2} (\partial \varphi)^2 + \frac{1}{2} (\partial \chi)^2 + \frac{1}{2} M^2 \chi^2 \right. \\
 \label{sinf}
&& \left. + \frac{g_0}{3!} \varphi^3 + \frac{\lambda_0}{4!} \varphi^4+ \frac{g_1}{2} \varphi^2 \chi + \frac{\lambda_1}{3!} \varphi^3 \chi \right] .
\end{eqnarray}
This is similar to the theory considered in \cite{Jackson:2010cw, Jackson:2011qg}, but we have added a $\varphi^3 \chi$ interaction to generate non-Gaussian interactions, and will now include self-interactions in our result \footnote{The fact that the interactions are tadpoles for $\chi$ is not an obstacle in applying our methods.}.  We will take the metric background to be exact de Sitter space,
\[ ds^2 = - dt^2 + a(t)^2 d{\bf x}^2 , \]
where ${\dot a}/a \equiv H = {\rm constant}$. 
\subsection{The In-In Formalism and Schwinger-Keldysh Basis}
In the Schwinger-Keldysh ``in-in" formalism, at some early time $t_{\rm in}$ we begin with a pure state $| {\rm in}(t_{\rm in}) \rangle$, then evolve the system for the bra- and ket-state separately until some late time $t$, when we evaluate the expectation value \cite{Calzetta:1986cq}:
\begin{eqnarray}
\label{inincorrelation}
\langle \mathcal O(t) \rangle &\equiv& \langle {\rm in}(t) | \mathcal O(t) | {\rm in} (t) \rangle \\
\nonumber
&=&  \langle {\rm in}(t_{\rm in} ) | e^{i \int _{t_{\rm in}} ^t  dt' H(t')} \mathcal O(t) e^{-i \int _{t_{\rm in}} ^t  dt'' H(t'')} | {\rm in} (t_{\rm in}) \rangle .
\end{eqnarray}
Traditionally, the in-state $| {\rm in} \rangle$ is taken to be the Bunch-Davies vacuum state \cite{Bunch:1978yq}, but this is not necessarily so.  Expanding cosmological backgrounds allow for a more general class of vacua, which can be heuristically considered to be excited states of inflaton fluctuations.  In the present context, we will find that integrating out high-energy physics generically results in corrections to the density matrix, which can be interpreted this way.

If we denote the fields representing the ``evolving'' ket to be $\{ \varphi_+, \chi_+ \}$ and those for the ``devolving'' bra to be $\{ \varphi_-, \chi_-  \}$, the in-in expectation value (\ref{inincorrelation}) can be computed from the action
\begin{equation}
\label{keldyshaction}
 \mathcal S \equiv S[\varphi_+, \chi_+] - S[\varphi_-, \chi_-] . 
 \end{equation}
together with the constraint that $\varphi_+(t)=\varphi_-(t)$.  It is then helpful to transform into the Keldysh basis,
\begin{eqnarray*}\
\nonumber
 {\bar \varphi} &\equiv& (\varphi_+ + \varphi_-)/2,  \hspace{0.5in} {\Phi} \equiv \varphi_+ - \varphi_-, \\
{\bar \chi} &\equiv& (\chi_+ + \chi_-)/2,  \hspace{0.5in} {\rm X} \equiv \chi_+ - \chi_-.
\end{eqnarray*}
In this basis the action (\ref{keldyshaction}) equals
\begin{eqnarray}
\nonumber
&& \hspace{-0.2in} \mathcal S [ {\bar \varphi}, {\Phi}, {\bar \chi},  {\rm X}] =  - \int d^3 {\bf x} dt \ a(t)^3 \left[ \partial {\bar \varphi} \partial \Phi + \partial {\bar \chi} \partial {\rm X} + M^2  {\bar \chi} {\rm X} \right. \\
\nonumber
&& \hspace{0.5in} + \frac{g_0}{2} \left( {\bar \varphi}^2 \Phi + \frac{1}{12} \Phi^3 \right) + \frac{\lambda_0}{3!} \left( {\bar \varphi}^3 \Phi + \frac{1}{4} {\bar \varphi} \Phi^3 \right) \\
\nonumber
&& \hspace{0.5in} \left. + g_1 {\bar \varphi} \Phi {\bar \chi} + \frac{g_1}{2} {\bar \varphi}^2 {\rm X}  + \frac{\lambda_1}{2} {\bar \varphi}^2 \Phi {\bar \chi} + \frac{\lambda_1}{3!} {\bar \varphi}^3 {\rm X} \right].
\end{eqnarray}
Fluctuation solutions are easiest written in the conformal time $\tau$,
\[ dt \equiv a(\tau) d \tau. \]
The free field solutions are then
\begin{eqnarray*}
U_{\bf k}(\tau) &=& \frac{H}{ \sqrt{2k^3}} (1 - ik \tau) e^{-ik \tau}, \\
V_{\bf k} (\tau) &\approx& \frac{1}{a(\tau)} \frac{\exp \left[ - i \int ^\tau_{\tau_{\rm in}} d \tau' \sqrt{  k^2 + \frac{M^2}{H^2 \tau^{\prime 2}}} \right]}{\sqrt{ 2} \left( k^2 + \frac{M^2}{H^2 \tau^2} \right)^{1/4} } 
\end{eqnarray*}
where for the latter we have used the WKB approximation which is always valid for $H/M \ll 1$. 

We will also need the Green's and Wightman functions in the Keldysh basis; see \cite{Jackson:2011qg} for an explanation of their interpretation.  Fourier transforming into comoving momentum, the retarded Green's function $G^{R}$ can be written in terms of the fluctuation solutions,
\begin{eqnarray}
\label{keldyshgreens}
G^R_{\bf k}(\tau_1,\tau_2) &\equiv& i \langle {\bar \varphi}_{\bf k} (\tau_1) \Phi_{\bf -k}(\tau_2) \rangle \\
\nonumber
&=& - 2 \theta(\tau_1-\tau_2) {\rm Im} \left[ U_{\bf k}(\tau_1) U^*_{\bf k}( \tau_2) \right] .
\end{eqnarray}
The advanced Green's function $G^A$ is then simply the time-reversal of this:
\[ G^A_{\bf k}(\tau_1,\tau_2) \equiv G^R_{\bf k}(\tau_2,\tau_1) . \]
A similar procedure applies for ${\bar \chi}$ using its corresponding retarded Green's function $\mathcal G^R$,
\begin{eqnarray}
\label{keldyshgreens}
\mathcal G^R_{\bf k}(\tau_1,\tau_2) &\equiv& i \langle {\bar \chi}_{\bf k} (\tau_1) {\rm X}_{\bf -k}(\tau_2) \rangle \\
\nonumber
&=& - 2 \theta(\tau_1-\tau_2) {\rm Im} \left[ V_{\bf k}(\tau_1) V^*_{\bf k}( \tau_2) \right].
\end{eqnarray}
The Wightman functions are
\begin{eqnarray*}
F_{\bf k}(\tau_1,\tau_2) &=& \langle {\bar \varphi}_{\bf k}(\tau_1) {\bar \varphi}_{\bf -k}(\tau_2) \rangle \\
&=& {\rm Re} \left[ U_{\bf k}(\tau_1) U^*_{\bf k}( \tau_2) \right] ,  \\
0 &=&\langle \Phi_{\bf k}(\tau_1) \Phi_{\bf -k}(\tau_2) \rangle, \\
\mathcal F_{\bf k}(\tau_1,\tau_2) &=& \langle {\bar \chi}_{\bf k}(\tau_1) {\bar \chi}_{\bf -k}(\tau_2) \rangle \\
&=& {\rm Re} \left[ V_{\bf k}(\tau_1) V^*_{\bf k}( \tau_2) \right] ,  \\
0 &=&\langle {\rm X}_{\bf k}(\tau_1) {\rm X}_{\bf -k}(\tau_2) \rangle.
 \end{eqnarray*}
We will now use these to compute higher-point correlation functions in the interacting theory.

\section{Bispectrum Corrections}
\subsection{Definitions}
A more precise definition of the bispectrum given by eq. (\ref{bdef}) is
\begin{eqnarray}
\label{ps}
&& \hspace{-0.5in}B_{\varphi} ({\bf k}_1, {\bf k}_2, {\bf k}_3) (2 \pi)^3  \delta^3 \left( {\bf k}_1+{\bf k}_2+{\bf k}_3 \right) \equiv \\
\nonumber
&& \hspace{-0.0in} \langle {\rm in} (0)  |  {\bar \varphi}_{{\bf k}_1} (0) {\bar \varphi}_{{\bf k}_2} (0)  {\bar \varphi}_{{\bf k}_3} (0) | {\rm in}(0) \rangle, 
\end{eqnarray}
where the time evolution of the in-state is given by
\[ | {\rm in}(\tau) \rangle = e^{- i \int^{\tau}_{\tau_{\rm in}} dt' \mathcal H(t') }| {\rm in}(\tau_{\rm in}) \rangle, \]
recalling that $\tau \rightarrow 0^-$ corresponds to future infinity.  From henceforth we will assume a Bunch-Davies in-state.  There are two types of contributions to $B_\varphi$: those arising from self-interactions of the $\varphi$ field, and those mediated by the heavy field $\chi$ coupled to $\varphi$.  Since they have qualitatively different behavior, we will consider each separately.
\subsection{Self Interactions}
\begin{figure}
\begin{center}
\hspace{-0.3in}
\parbox{40mm}{\includegraphics[scale=0.22]{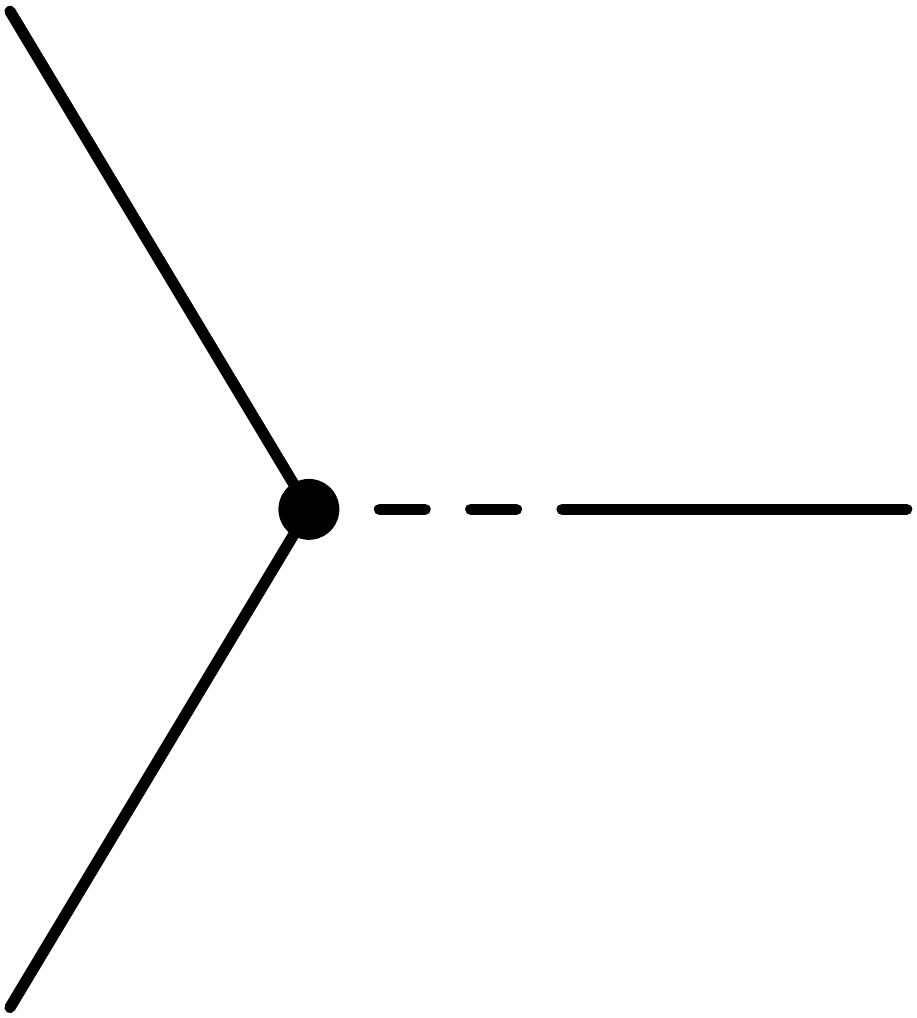} \hspace{-0.5in} }
\parbox{40mm}{\includegraphics[scale=0.22]{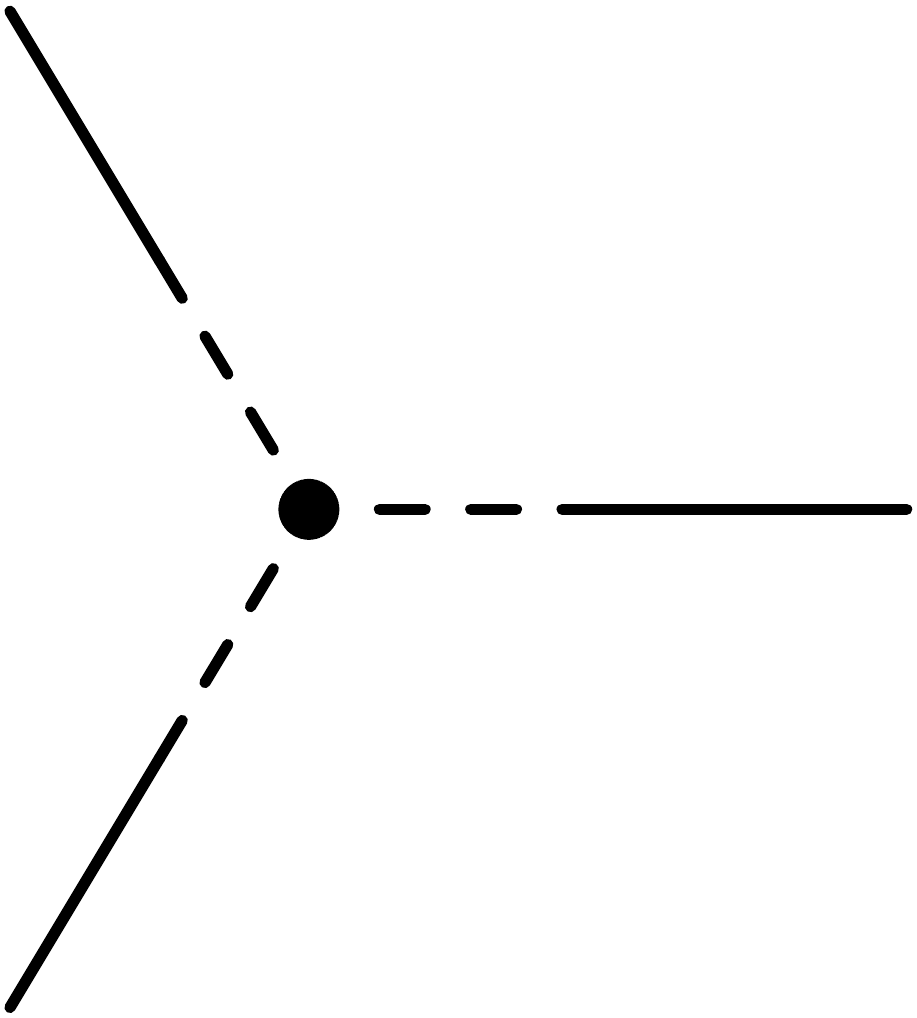} \hspace{-0.5in} } 
\caption{Lowest-order contributions to primordial bispectrum induced by the self-interaction of the light field.  Single solid lines indicate contractions of ${\bar \varphi}$, dashed single lines indicate those of $\Phi$.}
\label{nG_local_diags}
\end{center}
\end{figure}

\begin{figure}
\begin{center}
\hspace{-0.3in}
\parbox{40mm}{\includegraphics[scale=0.18]{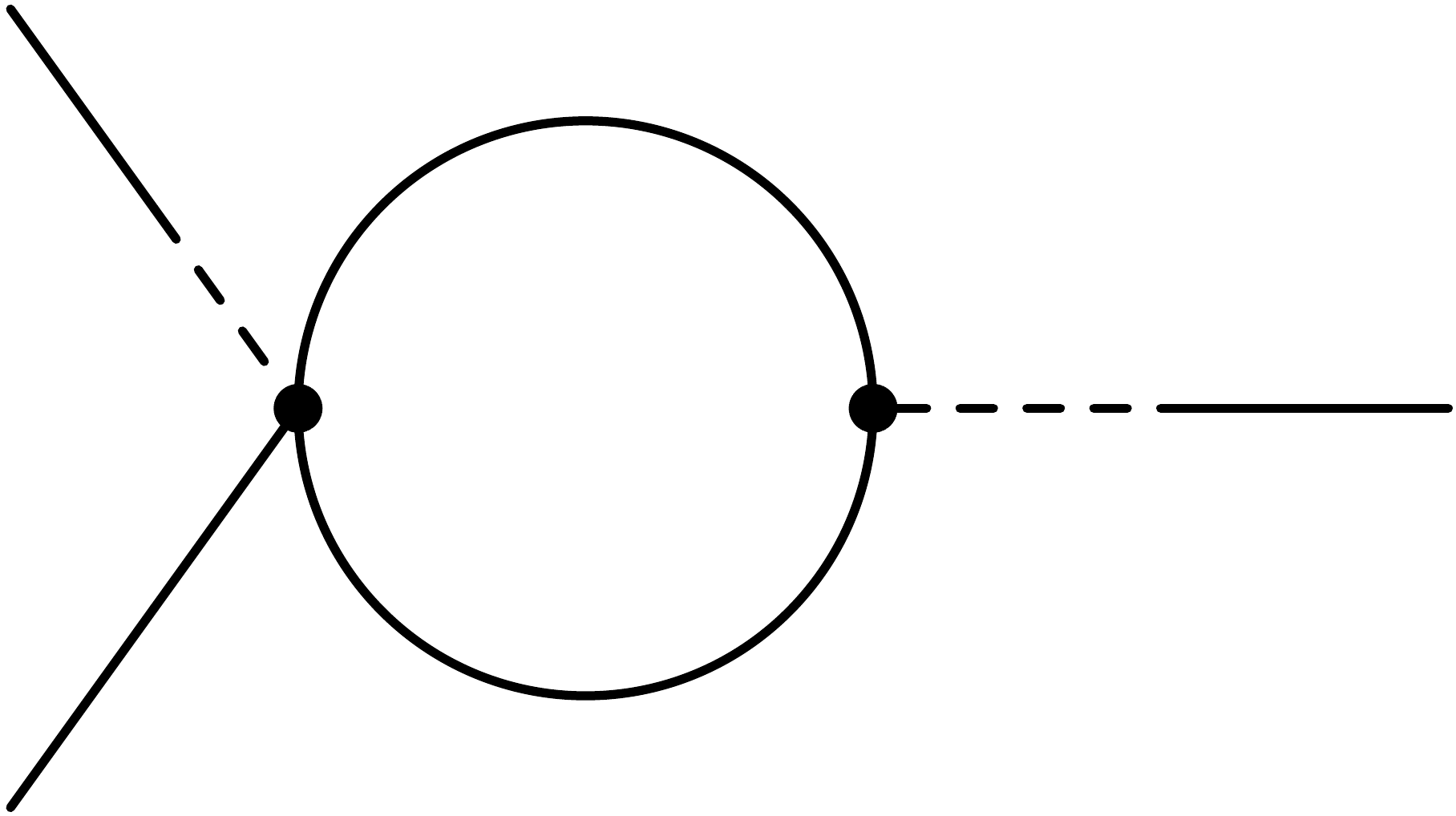} \hspace{-0.5in} }
\parbox{40mm}{\includegraphics[scale=0.22]{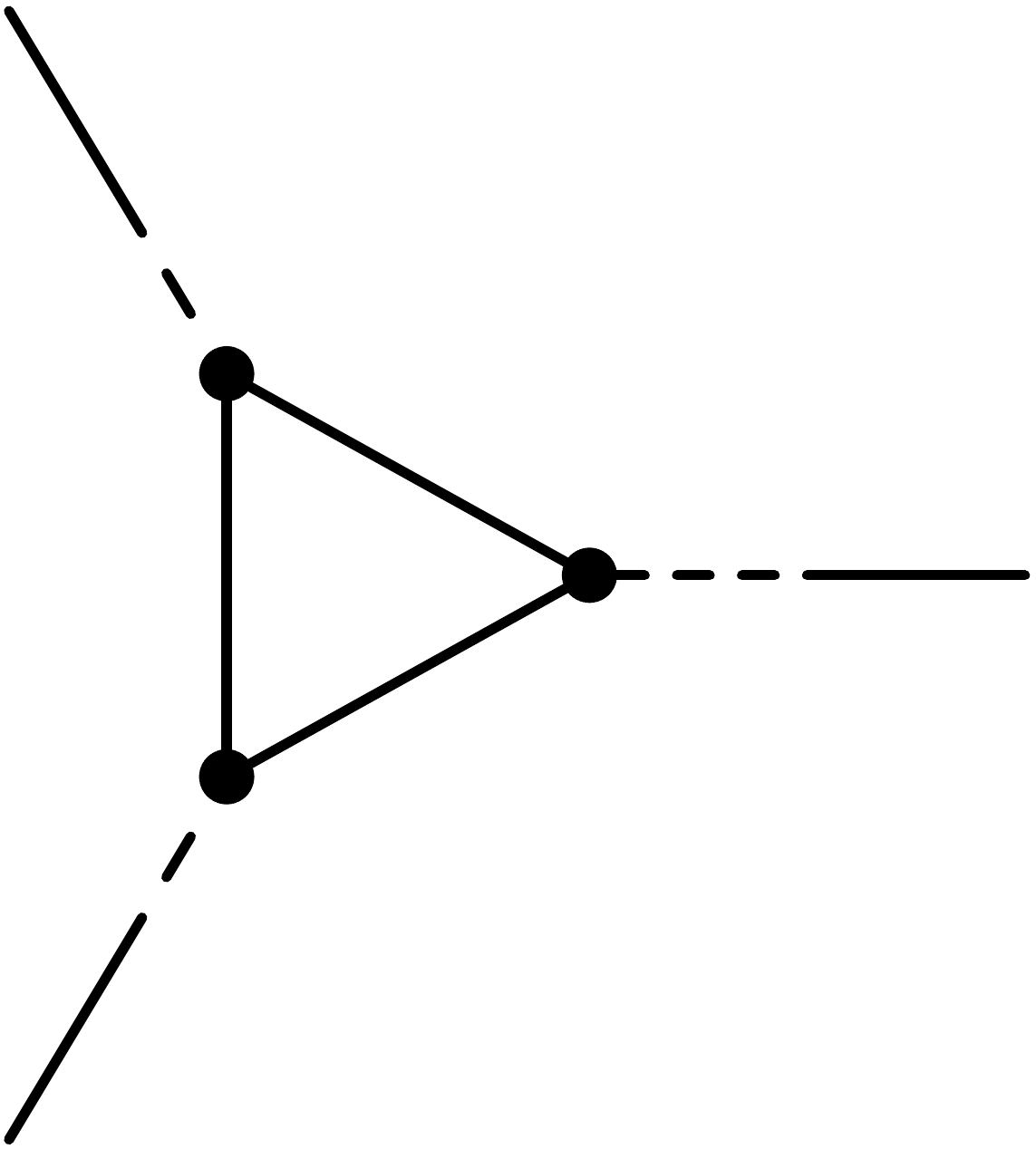} \hspace{-0.5in} } 
\caption{Higher-order bispectrum contribution induced by multiple self-interactions of the light field.  These can be neglected in slow-roll inflation due to the strong constraints on the couplings.}
\label{nG_local_diags_not}
\end{center}
\end{figure}

Let us first consider the simplest possible interaction: that arising from the $\varphi^3$ interaction.  In the Schwinger-Keldysh basis this produces two diagrams, shown in Figure~\ref{nG_local_diags}.  These are given by (we henceforth omit the momentum-conserving delta-function and just assume that ${\bf k}_1 + {\bf k}_2 + {\bf k}_3 = 0$):
\begin{eqnarray*}
B_\varphi^{\rm self} &=& \frac{(-ig_0)}{(2 \pi)^3} \int^0_{\tau_{\rm in}} d\tau \ a(\tau)^4  \\
&& \hspace{-0.5in} \times \left. \Big( [-iG^R_{{\bf k}_1} (0,\tau)] F_{{\bf k}_2}(0,\tau)  F_{{\bf k}_3}(0,\tau)+  \ {\rm permutations} \right. \\
&& \hspace{-0.5in} + \left. \frac{1}{4} [-iG^R_{{\bf k}_1} (0,\tau)][-i G^R_{{\bf k}_2}(0,\tau) ][-iG^R_{{\bf k}_3}(0,\tau)] \right).
\end{eqnarray*}
Writing out the Green and Wightman functions in terms of the functions $U_{\bf k}(\tau), U_{\bf k}^*(\tau)$, and using the fact that $U_{\bf k}(0) = U_{\bf k}^*(0) = H/\sqrt{2 k^3}$ yields
\begin{eqnarray*}
&& \hspace{-0.2in} B_\varphi^{\rm self} = \frac{i g_0 }{2 (2 \pi)^3 (2 k_1 k_2 k_3)^{3/2} H} \int _{\tau_{\rm in}}^0 \frac{d \tau}{\tau^4} \Big[ {U}_{{\bf k}_1}(\tau)  {U}_{{\bf k}_2}(\tau)  U_{{\bf k}_3}(\tau) \\
&& + {U}^*_{{\bf k}_1}(\tau)  {U}_{{\bf k}_2}(\tau)  U_{{\bf k}_3}(\tau) + {\rm permutations} - \ {\rm c.c.}  \Big] .
\end{eqnarray*}
The integrals will converge in the far past by slightly rotating the path of integration into the complex plane, so that $\tau_{\rm in} \rightarrow - \infty(1+ i\epsilon)$.  For the far future, we choose a cutoff at $\mu \rightarrow 0^-$, so that the most singular part of the resultant integrals are:
\begin{equation}
\label{gammaint}
 \int^\mu_{-\infty(1+ i\epsilon)} \frac{d \tau}{\tau^n} e^{- i k \tau} \sim - \frac{ \mu^{1-n}}{n-1} + \mathcal O\left( (k \mu ) ^{-n} \right).
 \end{equation}
Thus the late-time correlation is approximately
\[ B_\varphi^{\rm self} \approx \frac{g_0 \mu^{-3} H^2}{6(2 \pi k_1 k_2 k_3)^3} . \]
This correlation peaks when $k_i \approx 0$, corresponding to the squeezed shapes of the momentum triangle.  Note that this is \emph{not} quite the same local interaction studied by Komatsu and Spergel \cite{Komatsu:2001rj}, which introduce nonlinearities into the late-time gauge-independent parameter $\zeta$, whereas we deal with the field fluctuation $\varphi$ and integrate the interactions from the far past.  Defining the non-linearity parameter $f^{\rm self}_{\rm NL} \sim B^{\rm self}_{\varphi} /P^2_\varphi \sim k^{-3}$, the self interaction is seen to have a strong red scaling.  This may possibly be measurable in the near future \cite{Sefusatti:2009xu} using the leverage provided by combining the \emph{Planck} \cite{planck} and \emph{Euclid} \cite{euclid} datasets.  Employing two self interactions will produce loop corrections of order $B^{\rm self}_\varphi~\sim~g_0\lambda_0H^2$ and $B^{\rm self}_\varphi~\sim~g_0^3$ as shown in Figure \ref{nG_local_diags_not}, but slow-roll assumes that the these couplings are small and so these contributions are subleading.  

\subsection{High-Energy Correction A}
\begin{figure}
\begin{center}
\hspace{-0.3in}
\parbox{40mm}{\includegraphics[scale=0.18]{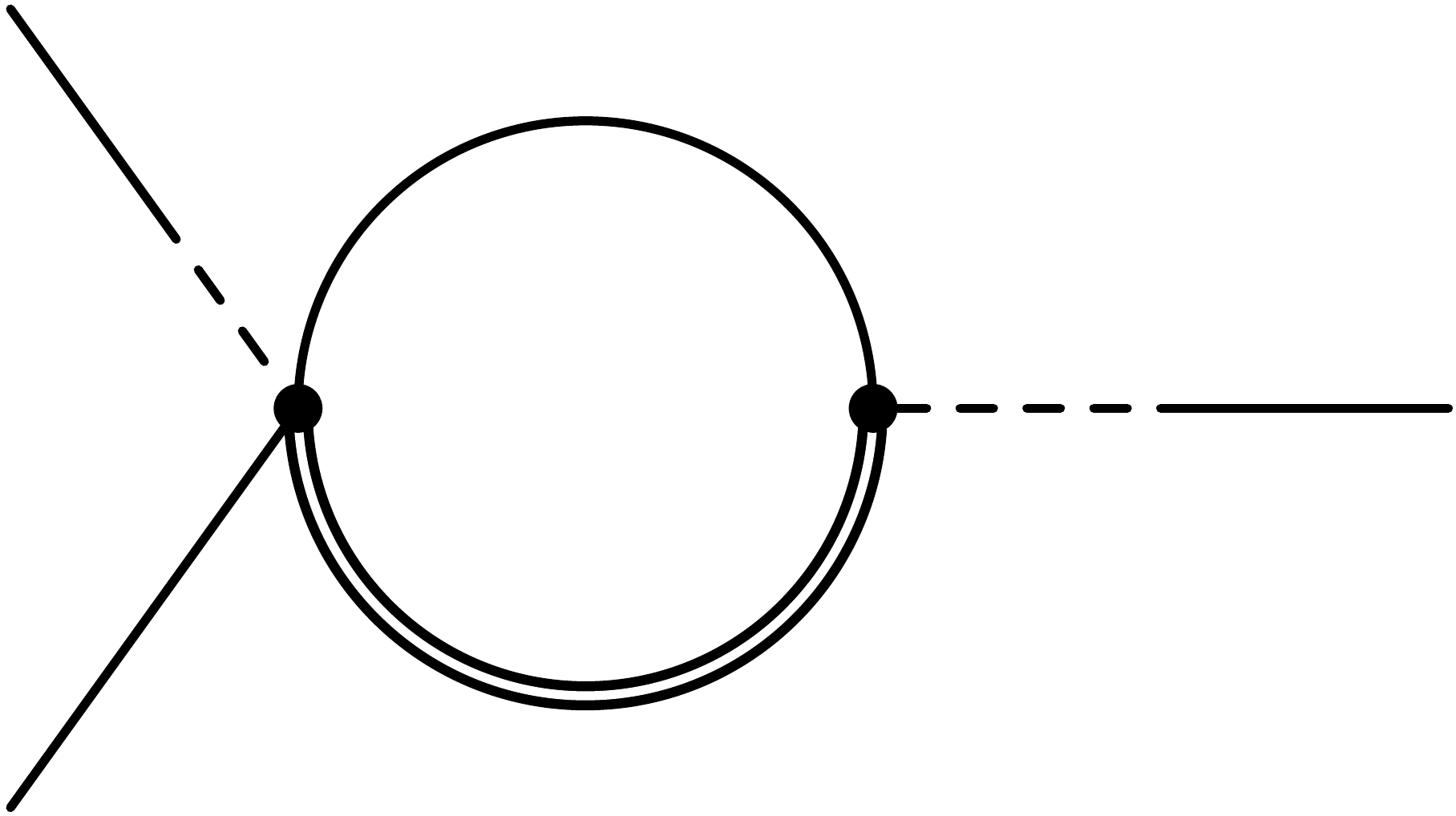} \hspace{-0.5in} \newline A}
\parbox{40mm}{\includegraphics[scale=0.18]{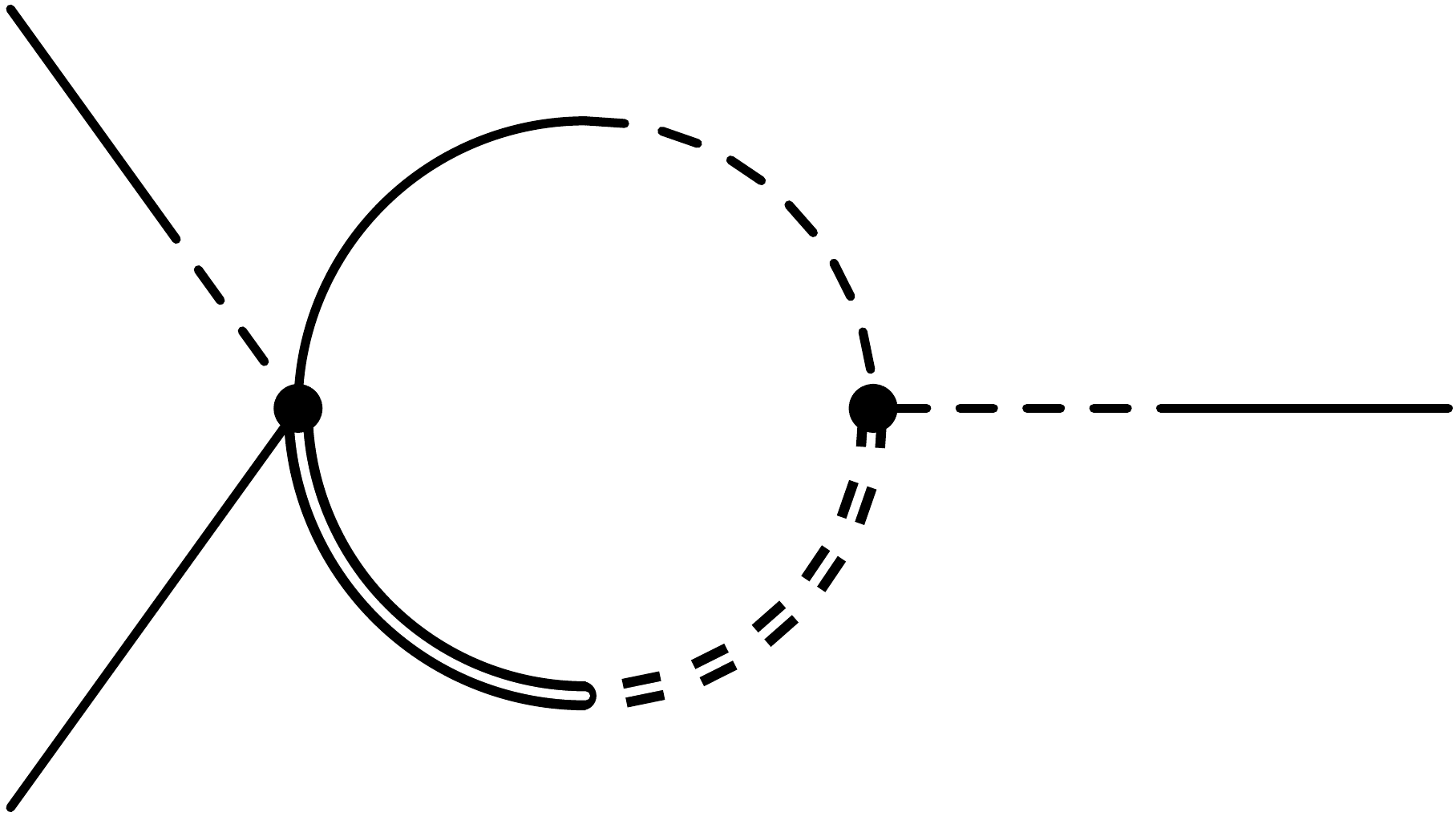} \hspace{-0.5in} \newline B} \\
\vspace{0.2in} \hspace{-0.3in}
\parbox{40mm}{\includegraphics[scale=0.18]{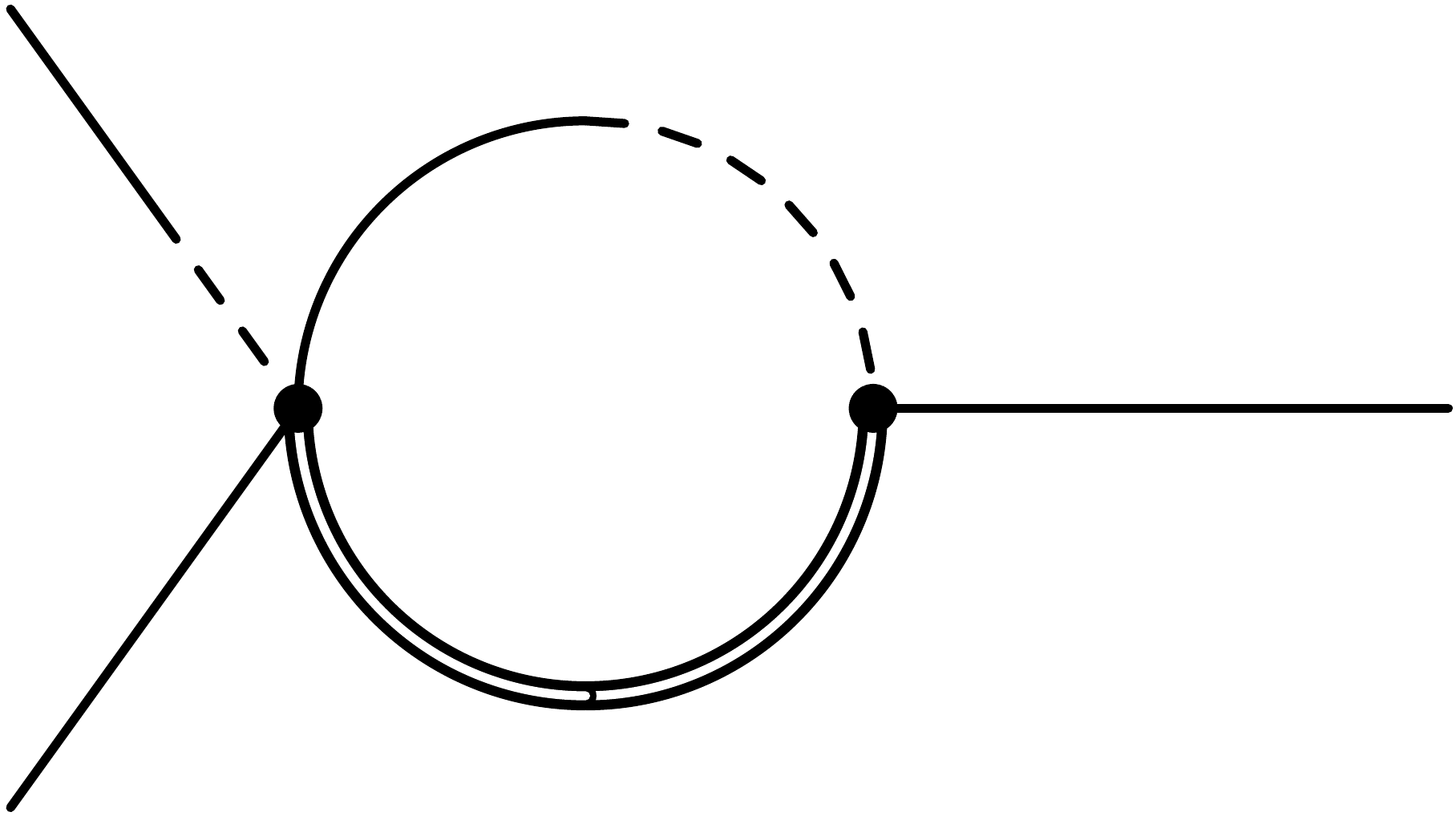} \hspace{-0.5in} \newline C} 
\parbox{40mm}{\includegraphics[scale=0.18]{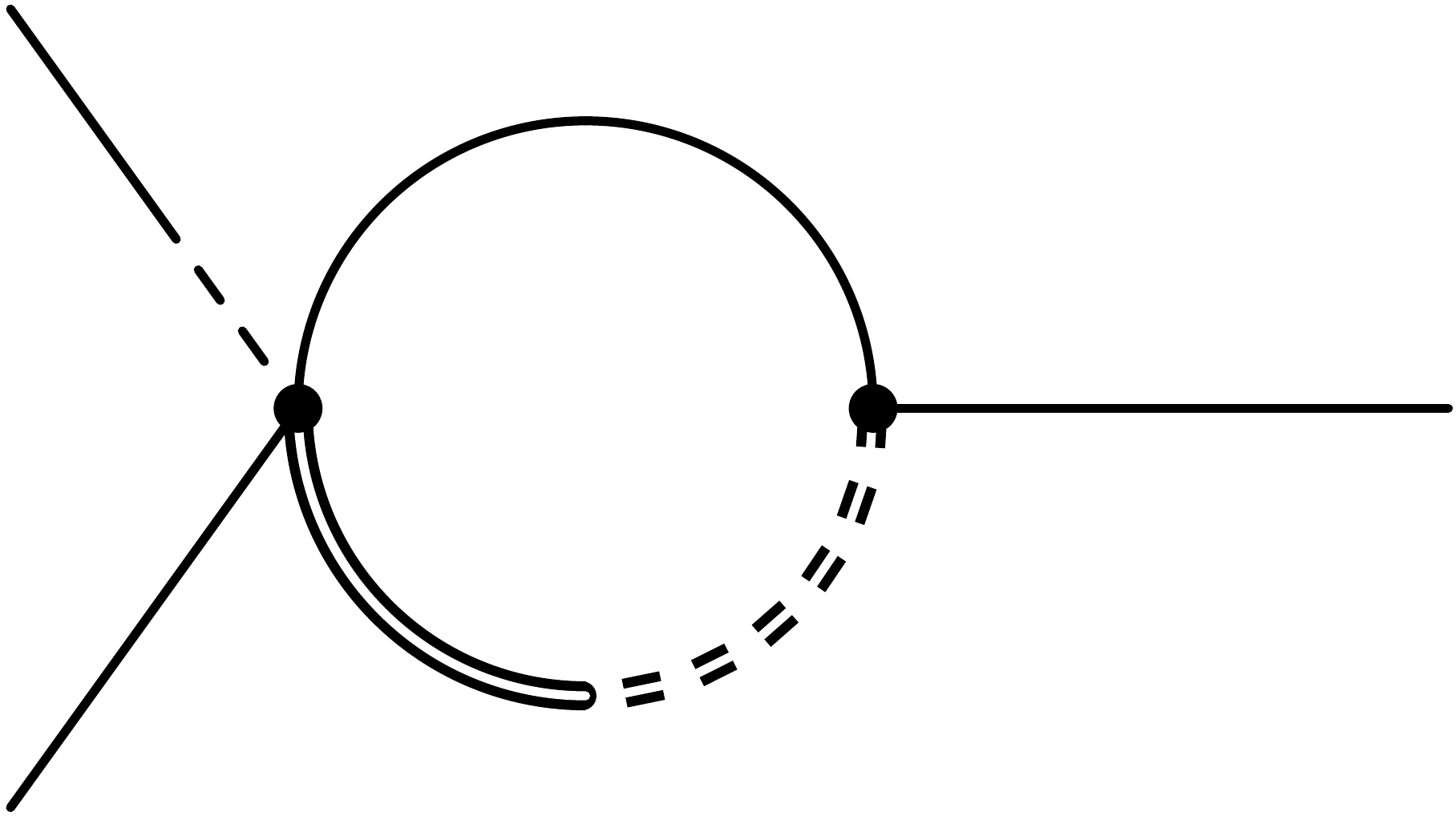} \hspace{-0.5in} \newline D}
\caption{Contributions to primordial non-Gaussianity induced by the heavy field.  The double lines indicate the heavy field components $\{ {\bar \chi},{\rm X} \}$, with notation analogous to Figure~\ref{nG_local_diags}.}
\label{nG_heavy_diags}
\end{center}
\end{figure}
We now turn to the corrections generated by high-energy physics.   To leading order in the heavy field, there are four such corrections, shown diagramatically in Figure~\ref{nG_heavy_diags}.  Since $\tau_2 \leq \tau_1$ we need only consider corrections containing $G^R(\tau_1,\tau_2)$ (as opposed to corrections containing $G^A(\tau_1,\tau_2)$).

The first correction is given by
\begin{eqnarray*}
B_\varphi^A &=& \frac{(-ig_1)(-i\lambda_1)}{(2 \pi)^3} \int^0_{\tau_{\rm in}} d\tau_1 \ a(\tau_1)^4 \int^0_{\tau_{\rm in}} d\tau_2 \ a(\tau_2)^4 \times \\
&& \hspace{-0.6in} \int \frac{d^3 {\bf q}}{(2 \pi)^3} \Big( [-iG^R_{{\bf k}_1} (0,\tau_1)] F_{{\bf k}_2}(0,\tau_1) F_{{\bf q}}(\tau_1,\tau_2) \times \\
&& \mathcal F_{ {\bf q}-{\bf k}_3}(\tau_1,\tau_2) [-iG^A_{{\bf k}_3} (\tau_2,0) ] + \ {\rm permutations} \Big).
\end{eqnarray*}
Writing out the Green and Wightman functions in terms of $U$'s and $V$'s, we see there are now several types of vertices, depending on the permutation of conjugation.  The first is
\begin{eqnarray*}
\mathcal A_1( {\bf k}_1,{\bf k}_2) &\equiv& \int _{\tau_{\rm in}}^0 d \tau \ a(\tau)^4 {U}_{{\bf k}_1}(\tau)  {U}_{{\bf k}_2}(\tau)  V^*_{-({\bf k}_1+{\bf k}_2)}(\tau) f(\tau) \\
&& \hspace{-0.5in} = - \frac{1}{2  \sqrt{2k_1^3 k_2^3} H } \int _{\tau_{\rm in}}^0 \frac{d \tau}{\tau^3}  \frac{  \left( 1-i k_1 \tau \right) \left( 1-i k_2 \tau \right) }{\left( | {{\bf k}_1+{\bf k}_2}|^2 + \frac{M^2}{H^2 \tau^2} \right)^{1/4} }f(\tau)  \\
&& \hspace{-0.8in} \times \exp \left[ - i (k_1 + k_2) \tau + i \int ^{\tau}_{\tau_{\rm in}} d \tau' \sqrt{ | {{\bf k}_1+{\bf k}_2}|^2 + \frac{M^2}{H^2 \tau'^2}} \right] ,
\end{eqnarray*}
where we have introduced a function $f(\tau)$ to account for any step-functions.  As first noted in \cite{Jackson:2010cw}, by rescaling $u \equiv H \tau / M$ the vertex $\mathcal A_1( {\bf k}_1,{\bf k}_2)$ admits a stationary phase approximation at the energy-conservation moment
\begin{equation}
\label{statphase}
 k_1+ k_2 = \sqrt{  | {\bf k}_1+{\bf k}_2|^2 + u_c^{-2}} . 
 \end{equation}
The solution to this defines the New Physics Hypersurface (NPH),
\[ u_{c}^{-1}= - \sqrt{2k_1k_2(1-\cos\theta)}, \hspace{0.3in} \cos\theta=\frac{ {\bf k}_1 \cdot {\bf k}_2}{k_1 k_2}. \]
Then to leading order in $H/M$ the amplitude is
\begin{eqnarray}
\nonumber
\mathcal A_1( {\bf k}_1,{\bf k}_2)  &\approx& - \frac{ \sqrt{\pi i} f(\tau_c) e^{-i \frac{M}{H} \sqrt{ | {\bf k}_1 + {\bf k}_2|^2  u_{\rm in}^2+ 1} }}{ 2\sqrt{k_1 k_2} \left[ 2 k_1 k_2 (1- \cos \theta ) \right]^{1/4} \sqrt{ HM}} \\
\label{a1}
&&\hspace{-0.5in} \times  \left( \frac{ {k_1 + k_2 + \sqrt{2k_1 k_2 (1- \cos \theta)}}}{\sqrt{ | {{\bf k}_1+{\bf k}_2}|^2 +  u_{\rm in}^{-2} } + |u_{\rm in}|^{-1}} \right) ^{-i \frac{M}{H} } .
\end{eqnarray}
The physics of this is clear. This diagram accounts for the threshold production/decay of heavy particles at high redshift in the early universe.  Note that in order to evaluate $f(\tau_c)$, one should use the step-function appropriately ``averaged" due to the Gaussian fluctuations:
\begin{equation}
\label{avstep}
 \theta(\tau) = \left\{ 
\begin{array}{cc} 
1 & {\rm if} \ \tau > 0, \\ 
\sfrac{1}{2} & {\rm if} \ \tau = 0, \\ 
0  & {\rm if} \ \tau < 0. \\ 
\end{array} \right. 
\end{equation}

The second possible vertex is identical to $\mathcal A_1$ but with one $U$ conjugated.  This has only imaginary-time saddlepoint solutions.  Since our $\tau$-integral is confined to the real axis we will never pass over this point in our integration, and so this amplitude will be suppressed as $\mathcal A_2 \sim {\rm erf} (\frac{M}{H}) \sim \frac{H}{M} e^{-(M/H)^2}$, allowing us to neglect such interactions.  Finally we consider $\mathcal A_3$ which has both $U$'s conjugated and so admits no saddlepoint solutions, and thus can also be neglected.   Note also that while there are actually two solutions (differing in sign), we limit ourselves to the solution for which $u_c < 0$, corresponding to the inflating phase (see \cite{Jackson:2011qg} for details on this). 

We can now perform a similar analysis of four-field interactions.  The first such vertex is
\begin{eqnarray*}
&& \hspace{-0.3in} \mathcal B_1 ( {\bf k}_1,{\bf k}_2, {\bf k}_3) \\
&\equiv&  \int _{\tau_{\rm in}}^0 d \tau \ a(\tau)^4 {U}_{{\bf k}_1}(\tau)  {U}_{{\bf k}_2}(\tau) {U}_{{\bf k}_3}(\tau) V^*_{-({\bf k}_1+{\bf k}_2+{\bf k}_3)}(\tau) f(\tau) \\
&=& - \frac{1}{4 \sqrt{k_1^3 k_2^3 k_3^3}  } \int^0_{\tau_{\rm in}} \frac{d\tau}{\tau^3}  \frac{  \left( 1-i k_1 \tau \right) \left( 1-i k_2 \tau \right) \left( 1-i k_3 \tau \right) }{\left(  | {\bf k}_1 + {\bf k}_2 +  {\bf k}_3|^2 + \frac{M^2}{H^2 \tau^2} \right)^{1/4}} \\
&& \hspace{-0.4in} \times \exp \left[  -i (k_1+k_2+k_3) \tau + i \int_{\tau_{\rm in}}^{\tau} d \tau' \sqrt{ | {\bf k}_1 + {\bf k}_2 +  {\bf k}_3|^2 + \frac{M^2}{H^2 \tau'^2} } \right] . 
\end{eqnarray*}
Using the same rescaling, the stationary phase is now at
\[ u_c^{-1} = - \sqrt{  (k_1+k_2+k_3)^2 - | {\bf k}_1 + {\bf k}_2 +  {\bf k}_3|^2 } \]
which always exists (that is, it is real).  To leading order, the amplitude is
\begin{eqnarray*}
 \mathcal B_1( {\bf k}_1,{\bf k}_2, {\bf k}_3) &\approx& \frac{ i \sqrt{\pi i M}  f(\tau_c) e^{-i \frac{M}{H} \sqrt{ | {\bf k}_1 + {\bf k}_2+ {\bf k}_3|^2 u_{\rm in}^2 + 1} } }{4 \sqrt{ H k_1 k_2 k_3 }} \\
&\times& \left[ (k_1+k_2+k_3)^2 - | {\bf k}_1 + {\bf k}_2 +  {\bf k}_3|^2\right]^{-3/4}  \\
&&\hspace{-1.2in} \times  \left( \frac{ {k_1 + k_2 + k_3 + \sqrt{  (k_1+k_2+k_3)^2 - | {\bf k}_1 + {\bf k}_2 +  {\bf k}_3|^2 } }}{\sqrt{ | {{\bf k}_1+{\bf k}_2} + {\bf k}_3|^2 +  u_{\rm in}^{-2} } + |u_{\rm in}|^{-1}} \right) ^{-i \frac{M}{H} } .
\end{eqnarray*}
A similar vertex, but with the third light field outgoing (we place a bar there to remind us of this), is:
\begin{eqnarray*}
&& \hspace{-0.3in} \mathcal B_2 ( {\bf k}_1,{\bf k}_2 | {\bf k}_3) \\
&\equiv&  \int _{\tau_{\rm in}}^0 d \tau \ a(\tau)^4 {U}_{{\bf k}_1}(\tau)  {U}_{{\bf k}_2}(\tau) {U}^*_{{\bf k}_3}(\tau) V^*_{-({\bf k}_1+{\bf k}_2+{\bf k}_3)}(\tau) f(\tau) \\
&=& - \frac{1}{4 \sqrt{k_1^3 k_2^3 k_3^3}  } \int^0_{\tau_{\rm in}} \frac{d\tau}{\tau^3}  \frac{  \left( 1-i k_1 \tau \right) \left( 1-i k_2 \tau \right) \left( 1+i k_3 \tau \right) }{\left(  | {\bf k}_1 + {\bf k}_2 +  {\bf k}_3|^2 + \frac{M^2}{H^2 \tau^2} \right)^{1/4}} \\
&& \hspace{-0.4in} \times \exp \left[  -i (k_1+k_2-k_3) \tau + i \int_{\tau_{\rm in}}^{\tau} d \tau' \sqrt{ | {\bf k}_1 + {\bf k}_2 +  {\bf k}_3|^2 + \frac{M^2}{H^2 \tau'^2} } \right] . 
\end{eqnarray*}
The stationary phase for this is at
\[ u_c^{-1} = - \sqrt{  (k_1+k_2-k_3)^2 - | {\bf k}_1 + {\bf k}_2 +  {\bf k}_3|^2 } . \]
This is \emph{not} always real, and one must take care.  To leading order, the amplitude is
\begin{eqnarray*}
\mathcal B_2( {\bf k}_1,{\bf k}_2 | {\bf k}_3) &\approx& \frac{ i \sqrt{\pi i M} f(\tau_c) e^{-i \frac{M}{H} \sqrt{ | {\bf k}_1 + {\bf k}_2+ {\bf k}_3|^2 u_{\rm in}^2 + 1} } }{4 \sqrt{ H k_1 k_2 k_3}} \\
&\times& \left[ (k_1+k_2-k_3)^2 - | {\bf k}_1 + {\bf k}_2 +  {\bf k}_3|^2\right]^{-3/4}  \\
&&\hspace{-1.2in} \times  \left( \frac{ {k_1 + k_2 - k_3 + \sqrt{  (k_1+k_2 -k_3)^2 - | {\bf k}_1 + {\bf k}_2 +  {\bf k}_3|^2 } }}{\sqrt{ | {{\bf k}_1+{\bf k}_2} + {\bf k}_3|^2 +  u_{\rm in}^{-2} } + |u_{\rm in}|^{-1}} \right) ^{-i \frac{M}{H} } .
\end{eqnarray*}

Based on this, it is easy to see that the amplitude is
\begin{eqnarray*}
&& \hspace{-0.3in} B^A_\varphi = \frac{g_1 \lambda_1}{8(2 \pi)^3}  \frac{H^3}{(2 k_1 k_2 k_3)^{3/2}} \int \frac{d^3 {\bf q}}{(2 \pi)^3}  \\
&& \hspace{-0.2in} \Big[   \mathcal A_1 ( {\bf q}, -{\bf k}_3) \mathcal B_1^* ( {\bf k}_2, {\bf q} , {\bf k}_1 ) + {\rm c.c.}Ê\\
&& \hspace{-0.2in} -  \mathcal A_1 ( {\bf q}, -{\bf k}_3) \mathcal B_2^* ( {\bf k}_2, {\bf q} | {\bf k}_1 )  +  \mathcal A_1 ( {\bf q},- {\bf k}_3) \mathcal B_2^* ( {\bf k}_1, {\bf q} | {\bf k}_2) + {\rm c.c.} \\
&& \hspace{2in} + \ {\rm permutations}  \Big].
\end{eqnarray*}
But the $k_1 \leftrightarrow k_2$ symmetry will cancel the $\mathcal B_2$-dependent terms, leaving only the $\mathcal B_1$-dependent terms.

Examining the loop integral in the expression above, the integrand contains a rapidly oscillating component due to the difference in vertex interaction times is
\begin{equation}
\label{phase}
\hspace{-0.1in} \left( \frac{ | {\bf k}_1 + {\bf k}_2 |  + q + \sqrt{ (  | {\bf k}_1 + {\bf k}_2 |  + q)^2 -  | {\bf k}_1 + {\bf k}_2 +  {\bf q}|^2 } }{  {k_1 + k_2 + q + \sqrt{  (k_1+k_2+q)^2 - | {\bf k}_1 + {\bf k}_2 +  {\bf q}|^2 } }} \right)^{-i \frac{M}{H} } .
\end{equation}
Given such rapid oscillations of this phase, this could potentially be evaluated using (another) stationary phase approximation.  However, one can easily convince themself that there exist no such stationary phase points except in the trivial case of $k_3 = | {\bf k}_1 + {\bf k}_2 | = k_1 + k_2$, meaning the phase is identically zero.  Thus, for all configurations except this, the amplitude is suppressed as $H/M$.

To evaluate the integral, we then expand near this point,
\begin{eqnarray*}
\kappa_{12} &\equiv& k_1 + k_2 - k_3,
 \end{eqnarray*}
and the oscillating component (\ref{phase}) simplifies to
\[ e^{i \frac{M}{H} \kappa_{12}/ \sqrt{2 q  k_3 (1- \cos \theta)}},\]
where $\theta$ is the angle between $- {\bf k}_3 = {\bf k}_1 + {\bf k}_2$ and ${\bf q}$.

Now we must do the integral over ${\bf q}$.  As done in earlier work  \cite{Jackson:2010cw, Jackson:2011qg}, we first use the azimuthal symmetry around ${\bf k}_3$ to simplify the 3d integral into just two variables,
\[ \int \frac{ d^3 {\bf q}}{(2 \pi)^3} \rightarrow \frac{1}{(2 \pi)^2} \int q^2 dq d(1-\cos \theta). \]
We wish to transform this into the $(u,E)$-basis given by
\begin{eqnarray}
\label{uEdefs}
u^{-1} &\equiv& -  \sqrt{2 k_3q(1-\cos \theta)}, \\
\nonumber
E \equiv M q |u| &=& M \sqrt{ \frac{q}{2k_3 (1-\cos \theta)} }.
\end{eqnarray}
These can be easily inverted as
\[ q = \frac{E}{M|u|} , \hspace{0.4in} 1- \cos \theta = \frac{M}{2 k_3 E |u|}, \]
so that the ${\bf q}$-integral then transforms as
\begin{equation}  
\label{qtrans}
\hspace{-0.0in} \int \frac{d ^3 \bf q}{(2 \pi)^3} \rightarrow - \frac{1}{(2 \pi)^2 M^2 k_3} \int \frac{du E d E}{u^5 } . 
\end{equation}
The integrand transforms as
\begin{equation}
\label{abstar}
 \mathcal A_1 ( {\bf q}, -{\bf k}_3) \mathcal B_1^* ( {\bf k}_2, {\bf q} , {\bf k}_1 ) \rightarrow \frac{\pi |u|^3 M e^{- i \frac{M}{H} \kappa_{12} u } }{8 H E \sqrt{k_1 k_2 k_3} } + \mathcal O(\kappa_{12}) .
 \end{equation}

Noting that $E$ is a physical rather than a comoving scale, we can now easily place limits on the region of integration in terms of the physical cutoff $\Lambda$.  Recalling that at the moment of interaction the energies of the light fields will sum to the energy of the heavy field, and so it is this latter quantity that we must place the bound of $\Lambda$ on.  The minimal energy bound comes from the geometrical constraint $\cos \theta \geq -1$.  The bounds on the loop energy are then:
\begin{eqnarray*}
\frac{M}{4 k_3 |u|} \leq & E & \leq \Lambda - k_3 | u| M, \\
-\frac{\Lambda + \sqrt{\Lambda^2 - M^2}}{2 M k_3} \leq &u& \leq -\frac{\Lambda - \sqrt{\Lambda^2 - M^2}}{2 M k_3}.
\end{eqnarray*}

\begin{figure}
\begin{center}
\includegraphics[scale=0.8]{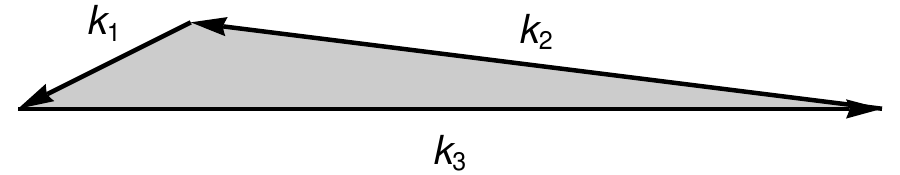}
\caption{Example of an `elongated' bispectrum momentum-vector triangle which can be produced from high-energy physics, where $k_1 + k_2 \approx k_3$.}
\label{bispect_HE}
\end{center}
\end{figure}

The measure (\ref{qtrans}) and integrand (\ref{abstar}) combine to produce an $E$-integral which is trivial, 
\[ \int^{\Lambda - k_3 | u| M}_\frac{M}{4 k_3 |u|} dE =  \Lambda - k_3 | u| M - \frac{M}{4 k_3 |u|}. \]
Now including the factor of $u^{-2}$ from the measure and vertex evaluations (\ref{abstar}) we can well-approximate the remaining integral over $u$ as Gaussian near the peak at $u_0 \approx - 3 M/8 \Lambda k_{3}$,
\begin{eqnarray*}
&& \hspace{-0.2in} |u|^{-2} \left[ \Lambda - k_3 | u| M - \frac{M}{4 k_3 |u|} \right] \\
 &\approx& \frac{\Lambda}{3} \left( \frac{8 \Lambda k_3 }{3 M} \right)^{2} \exp \left[ - \frac{2^7}{6} \left( \frac{\Lambda k_3 }{M} \right)^{2} (u-u_0)^2\right] + \mathcal O \left( \frac{M}{\Lambda} \right).
 \end{eqnarray*}
Extending the domain of integration to $u \in (-\infty, \infty)$, the integral over $u$ can then be easily performed to yield another Gaussian for $\kappa_{12}$.  The exponential will cut off anything beyond $\kappa_{12} \approx 2^4 k_3 \Lambda H / \sqrt{3} M^2$, by which time the phase will have completed about half a cycle.  Thus we can approximate this as a delta-function, 
\begin{eqnarray*}
&& \hspace{-0.3in} \int_{- \infty}^{\infty} du \ e^{ -i \frac{M}{H} \kappa_{12} u -  \frac{2^7}{6} \left( \frac{\Lambda k_3 }{M} \right)^{2} (u-u_0)^2 } + {\rm c.c.} \\
&\approx& \sqrt{ \frac{ 6 \pi}{2^5}}  \left( \frac{M}{\Lambda k_3 } \right) e^{ - i \frac{M}{H} \kappa_{12} u_0 - \frac{3}{2^8} \left( \frac{M^2  \kappa_{12}}{H \Lambda k_3} \right)^2 } + {\rm c.c.} \\
&\approx& \frac{4 \pi H}{M} \delta(\kappa_{12}) .
\end{eqnarray*}
An important caveat is that corrections to this approximation will be of order $\Lambda H / M^2$, necessitating that this be a small quantity.  This is the same restriction one obtains for using the plane wave approximation $U_{\bf k} \approx \frac{H \tau}{\sqrt{2 k}} e^{-i k \tau}$ for all interactions and so is already implicitly assumed. The final answer is then
\begin{equation}
\label{ba}
 B^A_\varphi \approx \frac{\pi g_1 \lambda_1 H^3 \Lambda^3 }{27  \sqrt{2}   (2 \pi)^4 M^4 k_1^2 k_2^2 k_3} \delta( k_1 + k_2 - k_3) +  {\rm perms.}
 \end{equation}
The requirement that $k_1 + k_2 \approx  k_3$ implies that the momentum-vector diagram approximates an elongated triangle, as shown in Figure~\ref{bispect_HE}.  The fact that modifications to the initial state produce the elongated type of non-Gaussianity was anticipated by \cite{Holman:2007na, Meerburg:2009ys, Meerburg:2009fi}; here we see how this originates from fundamental high-energy physics.  The factor of $(\Lambda/M)^3$ can be absorbed into the bare couplings $g_1, \lambda_1$.  Since $g_1 \sim~H, \lambda_1 \sim 1, \Lambda/M \sim 1$, the correction (\ref{ba}) will then scale as $B^A_\varphi \sim H/M$.  The nonlinearity parameter scaling as $f^{\rm A}_{\rm NL} \sim B^{\rm A}_{\varphi}/P^2_\varphi$ is scale-invariant, so $n^A_{\rm NG} \equiv d \ln f^A_{\rm NL}/ d \ln k = 0$.  

\subsection{High-Energy Correction B}
The next diagram is very similar,
\begin{eqnarray*}
B_ \varphi ^B &=& \frac{(-ig_1)(-i\lambda_1) }{(2 \pi)^3} \int^0_{\tau_{\rm in}} d\tau_1 \ a(\tau_1)^4 \int^0_{\tau_{\rm in}} d\tau_2 \ a(\tau_2)^4 \times \\
&& \hspace{-0.4in} \int \frac{d^3 {\bf q}}{(2 \pi)^3} \Big([ -i G^R_{{\bf k}_1} (0,\tau_1)] F_{{\bf k}_2}(0,\tau_1) [-i G^R_{{\bf q}}(\tau_1,\tau_2) ] \times \\
&& [-i \mathcal G^{R}_{ {\bf q}-{\bf k}_3}(\tau_1,\tau_2) ][ -i G^A_{{\bf k}_3} (\tau_2,0) ]  + \ {\rm permutations} \Big).
\end{eqnarray*}
We do not need to explicitly evaluate this, however, because in the stationary phase approximation this is easily seen to be identical to diagram A but with a minus sign and the Heaviside function $\theta(\tau_1-\tau_2)$.  This allows us to integrate over only half the fluctuations in $\tau_1,\tau_2$ and hence we get
\[ B_ \varphi ^B = - \frac{1}{2} B_ \varphi ^A. \]

\subsection{High-Energy Corrections C and D}
There are two additional diagrams from high-energy interactions,
\begin{eqnarray*}
B_ \varphi ^C &=& \frac{(-ig_1)(-i\lambda_1)}{(2 \pi)^3} \int^0_{\tau_{\rm in}} d\tau_1 \ a(\tau_1)^4 \int^0_{\tau_{\rm in}} d\tau_2 \ a(\tau_2)^4 \times \\
&& \hspace{-0.4in} \int \frac{d^3 {\bf q}}{(2 \pi)^3} \Big( [-iG^R_{{\bf k}_1} (0,\tau_1)] F_{{\bf k}_2}(0,\tau_1) [-i G^R_{{\bf q}}(\tau_1,\tau_2)] \times \\
&&  \mathcal F_{ {\bf q}-{\bf k}_3}(\tau_1,\tau_2) F_{{\bf k}_3} (\tau_2,0) + {\rm permutations} \Big), \\
B_\varphi^D &=& \frac{(-ig_1)(-i\lambda_1)}{(2 \pi)^3} \int^0_{\tau_{\rm in}} d\tau_1 \ a(\tau_1)^4 \int^0_{\tau_{\rm in}} d\tau_2 \ a(\tau_2)^4 \times \\
&& \hspace{-0.4in} \int \frac{d^3 {\bf q}}{(2 \pi)^3} \Big( [-i G^R_{{\bf k}_1} (0,\tau_1) ] F_{{\bf k}_2}(0,\tau_1) F_{{\bf q}}(\tau_1,\tau_2) \times \\
&& [-i \mathcal G^R_{ {\bf q}-{\bf k}_3}(\tau_1,\tau_2) ] F_{{\bf k}_3} (\tau_2,0) + {\rm permutations} \Big) .
\end{eqnarray*}
This time, the stationary phase approximation implies the total cancellation of these terms against each other.  As first observed in the power spectrum calculation \cite{Jackson:2010cw}, this is a general theme: the leading-order corrections arising from the dynamics effective action will cancel, and the dominant corrections by power counting arise entirely from the density matrix.
\subsection{Total Bispectrum}
\begin{figure}
\begin{center}
\hspace{-0.2in} \parbox{18mm}{\includegraphics[scale=0.18]{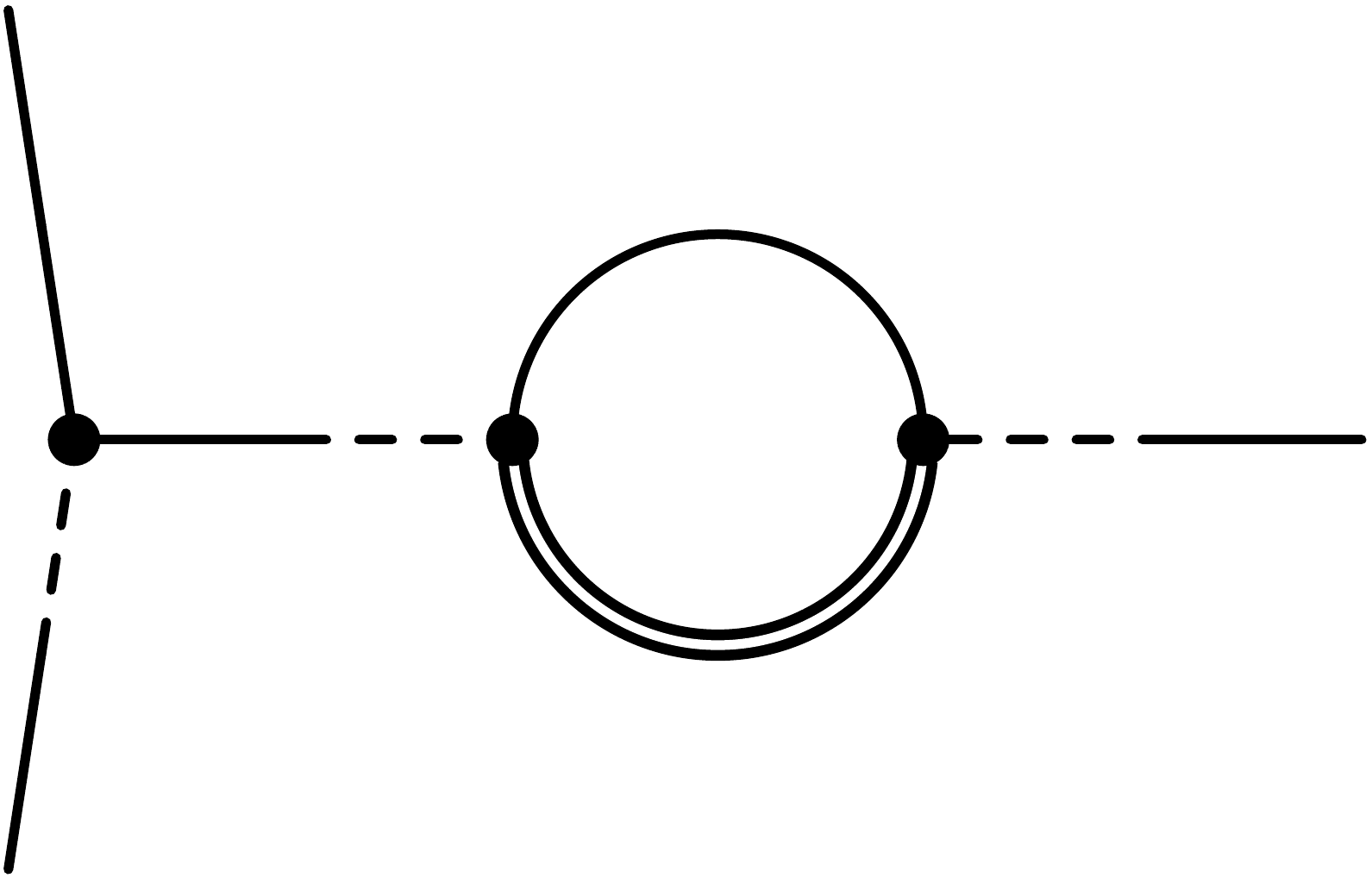}} \hspace{0.8in}
\parbox{16mm}{\includegraphics[scale=0.18]{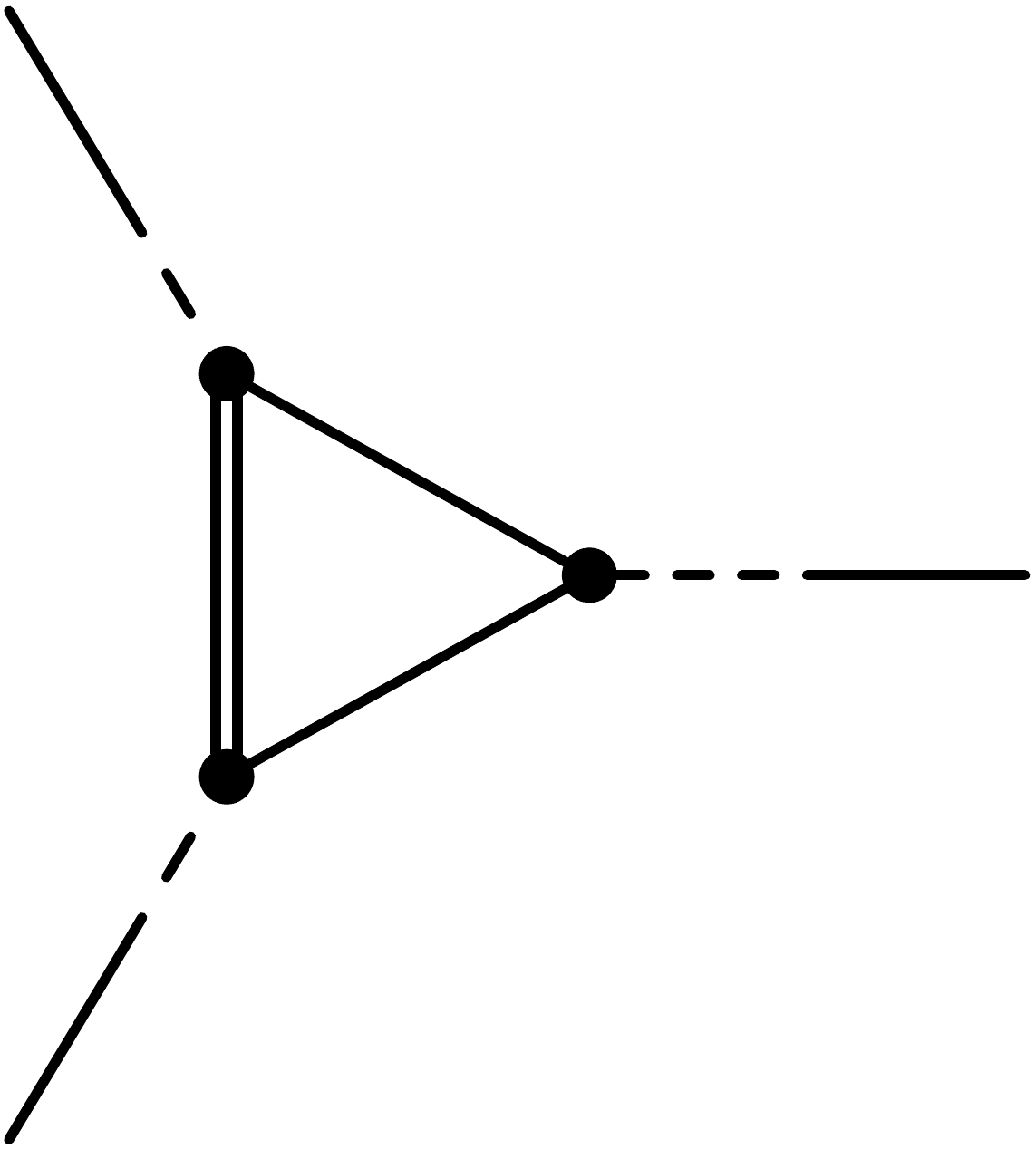}}
\caption{Examples of 3-point diagrams which are model-dependent, and so not considered here.}
\label{not_nG_diags}
\end{center}
\end{figure}
The complete bispectrum correction is then
\[ B_ \varphi = B_ \varphi ^{\rm self}+B_ \varphi ^A + B_ \varphi ^B.  \]
The high-energy physics produces a scale-invariant elongated shape of bispectrum corrections.  

There are a few other diagrams involving high-energy interactions shown in Figure~\ref{not_nG_diags}, but they all involve a local or derivative coupling (strongly constrained by slow-roll), as well as a factor of $H/M$.  Thus, these model-dependent terms will be subleading to the diagrams computed here.  In non-slow roll models local terms may significantly contribute and in fact correspond precisely to the terms discussed in \cite{Meerburg:2009fi}.  The vacuum modification should also enhance local interactions, as first found in \cite{Ganc:2011dy, Chialva:2011hc}.  Higher-derivative actions (for example, DBI \cite{Alishahiha:2004eh} or more generally \cite{Chen:2006nt}) favor large equilateral bispectra \cite{Creminelli:2003iq}, although will likely contain new features due to the high-energy physics. 

\section{Trispectrum Corrections}
The 4-point correlation, or trispectrum, is defined analogously to the bispectrum:
\begin{eqnarray}
\nonumber
&&\hspace{-0.5in} T_\varphi({\bf k}_1,{\bf k}_2,{\bf k}_3,{\bf k}_4) (2 \pi)^3 \delta^3 \left( {\bf k}_1+{\bf k}_2+{\bf k}_3 +{\bf k}_4 \right) \\
&\equiv& \langle {\rm in}(0) | \varphi_{{\bf k}_1} (0) \varphi_{{\bf k}_2} (0) \varphi_{{\bf k}_3} (0)  \varphi_{{\bf k}_4} (0) |  {\rm in}(0) \rangle  .
\end{eqnarray}
It was suggested in \cite{Jackson:2011qg} that highly oscillatory corrections to power spectrum may be invisible, yet show up in the trispectrum as enhanced variance. This is because the dominant contribution to the trispectrum is the power spectrum-squared,
\[ T_\varphi \sim \delta^3 ( {\bf k}_1 - {\bf k}_2 ) \delta^3 ( {\bf k}_3 - {\bf k}_4 ) P_\varphi (k_1) P_\varphi (k_3) + {\rm permutations} .\]
A modification in $P_\varphi$ of order $H/M$ would then produce a change of order $2 P_\varphi H/M$ in $T_\varphi$.  Below we will evaluate the corrections to $T_\varphi$, due to self- and high-energy interactions.  We will see that this is not true.
\subsection{Self Interactions}
\begin{figure}
\begin{center}
\hspace{-0.0in}
\parbox{35mm}{\includegraphics[scale=0.18]{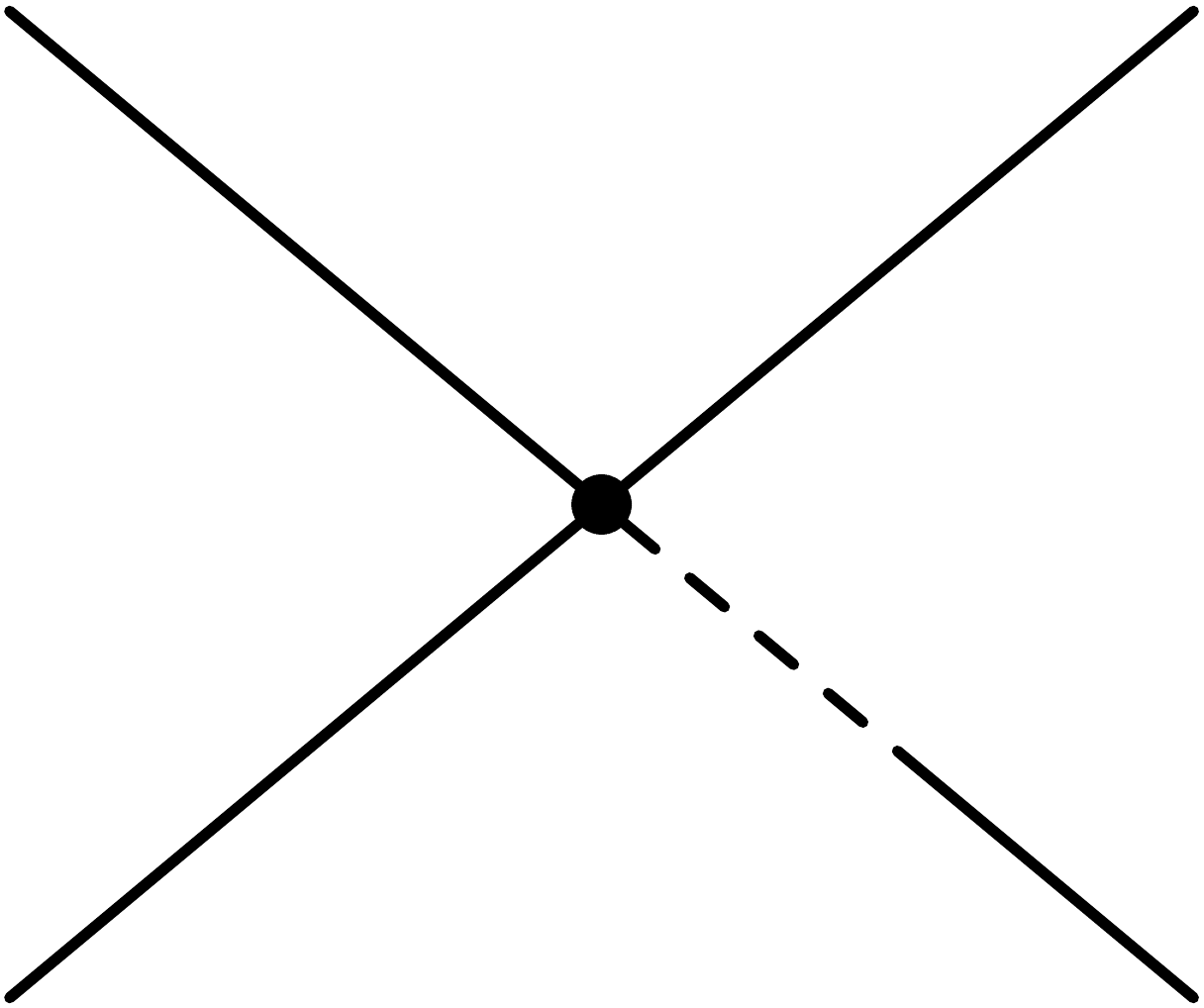}  }
\parbox{35mm}{\includegraphics[scale=0.18]{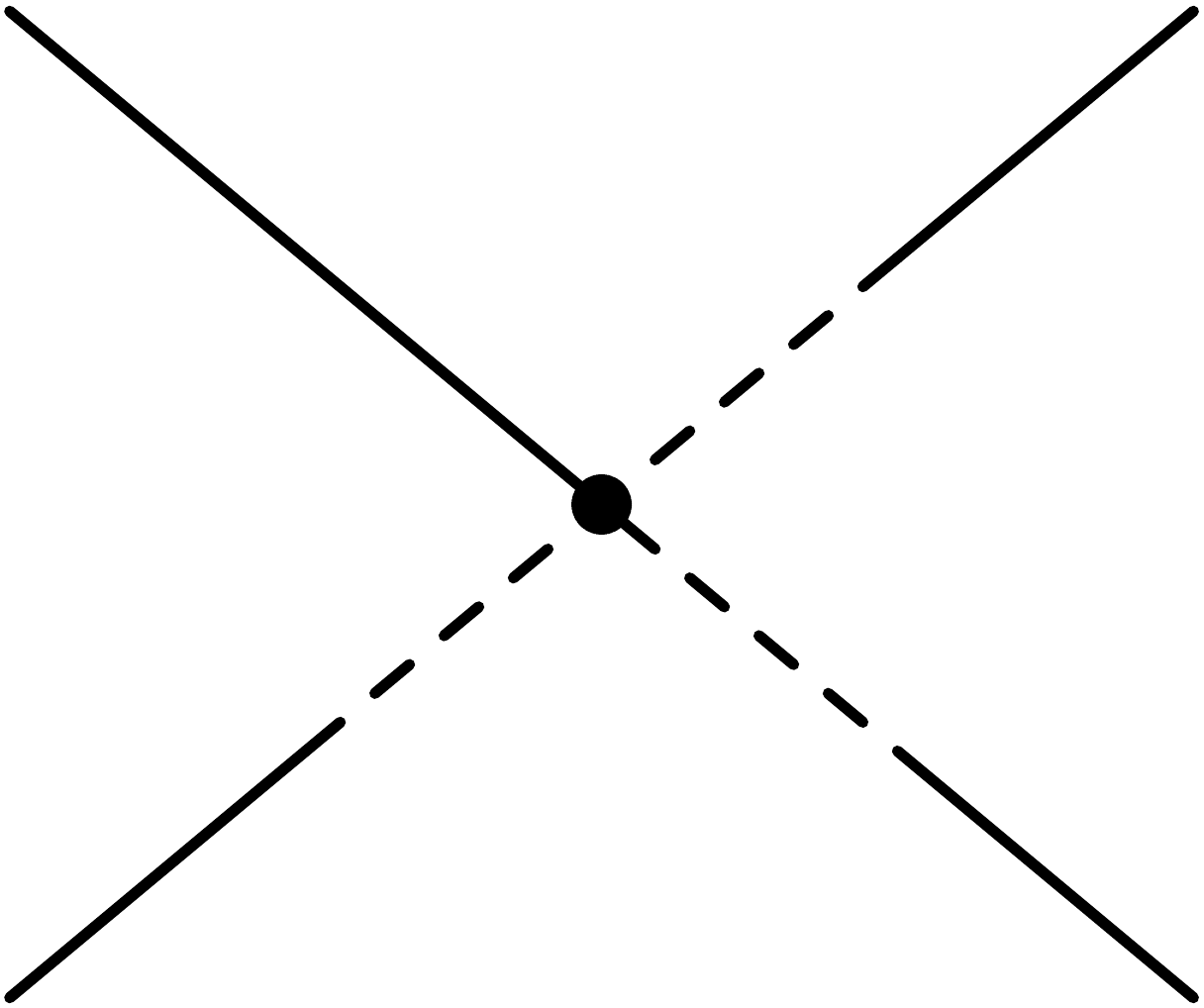} } 
\caption{Contributions to the trispectrum from self-interactions.  There are also $\mathcal O(g_0^2)$ and $\mathcal O(\lambda_0^2)$ contributions which we can neglect in slow-roll inflation due to the strong constraints on the couplings.}
\label{4pt_local_diags}
\end{center}
\end{figure}
The first non-factorizable contribution will be from the self-interaction as shown in Figure \ref{4pt_local_diags}, again using a late-time cutoff $\mu \rightarrow 0^-$ and the most singular part of the exponential integral given by (\ref{gammaint}),
\begin{eqnarray*}
T_ \varphi ^{\rm self} &=& \frac{(-i \lambda_0)}{(2 \pi)^3} \int^{\mu}_{-\infty(1+i \epsilon)} d\tau \ a(\tau)^4  \times \\
&& \hspace{-0.6in} \Big( F_{{\bf k}_1}(0,\tau) F_{{\bf k}_2}(0,\tau) F_{{\bf k}_3}(0,\tau) [-i G^R_{{\bf k}_4} (0,\tau)] + \ {\rm permutations}  \\
&&  \hspace{-0.6in} + F_{{\bf k}_1}(0,\tau) [-iG^R_{{\bf k}_2}(0,\tau)][-i G^R_{{\bf k}_3}(0,\tau) ][-iG^R_{{\bf k}_4} (0,\tau)] \\
&& + \ {\rm permutations} \Big) \\
&& \hspace{-0.5in} = \frac{i \lambda_0 }{8(2 \pi)^3(k_1 k_2 k_3 k_4)^{3/2}} \int^{\mu}_{-\infty(1+i \epsilon)} \frac{d \tau}{\tau^4} \times \\
&& \hspace{-0.3in}  \Big[ U^*_{{\bf k}_1} (\tau) U_{{\bf k}_2} (\tau) U_{{\bf k}_3} (\tau) U_{{\bf k}_4} (\tau)+ {\rm permutations} - {\rm c.c.} \Big] \\
&&  \hspace{-0.5in} \approx \frac{ \lambda_0 \mu^{-3} H^4 }{24(2 \pi k_1 k_2 k_3 k_4)^{3}}.
\end{eqnarray*}
Such local trispectrum correlation functions clearly peaks at $k_i \approx 0$, and has been studied in \cite{Hu:2001fa}.  This has a non-linearity parameter $g_{\rm NL}^{\rm self} \sim T^{\rm self}_\varphi / P^3_{\rm \varphi} \sim k^{-3}$ scaling.  There will also be corrections of order $T_\varphi \sim g_0^2 H^2$ and $T_\varphi \sim \lambda_0^2 H^4$, which are again presumed negligible from slow-roll constraints.  

\subsection{Tree-Level High-Energy Corrections}
\begin{figure}
\begin{center}
\hspace{-0.0in}
\parbox{35mm}{\includegraphics[scale=0.18]{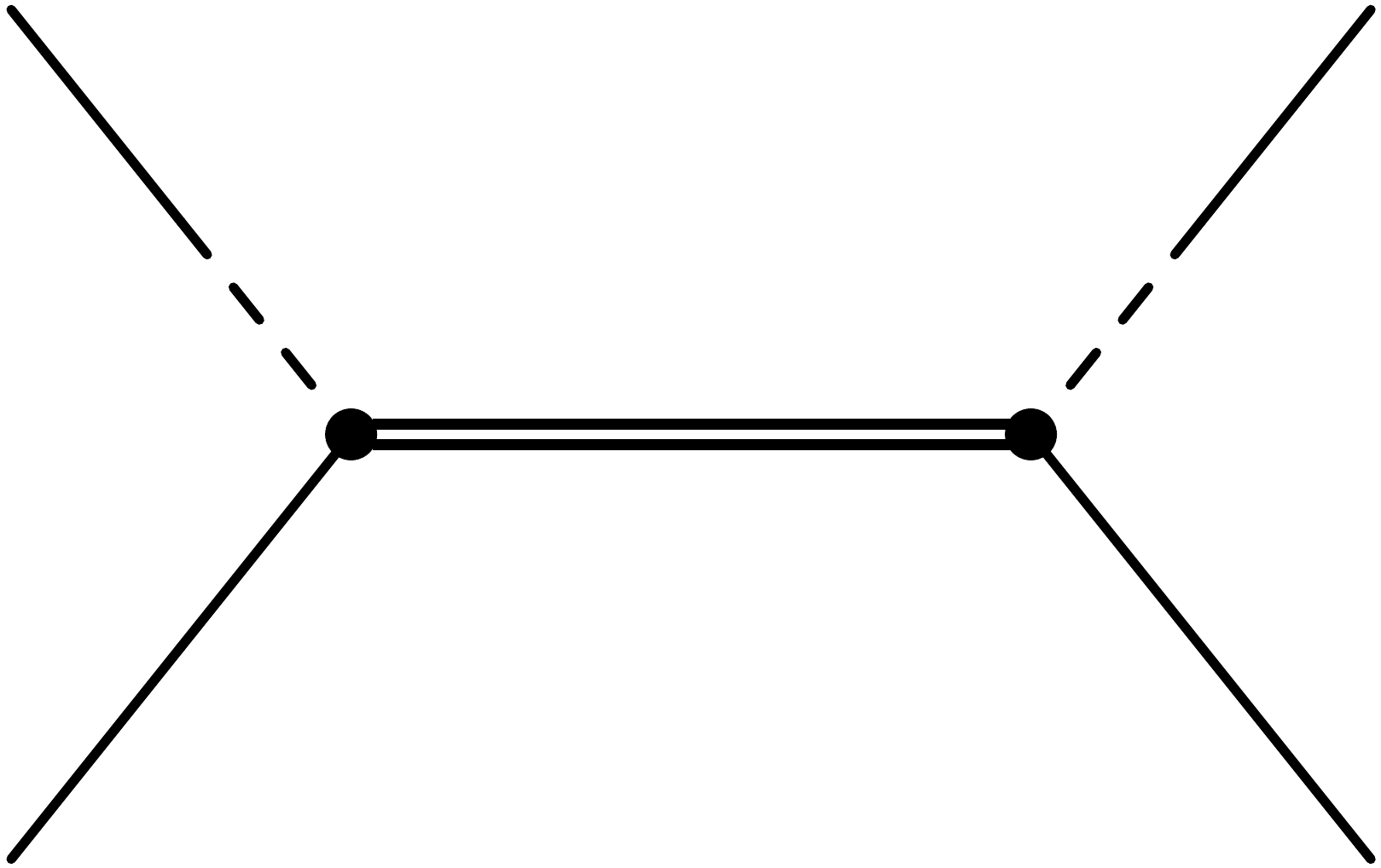} \hspace{-0.1in} \newline A} \\
\vspace{0.2in} \hspace{0.05in}
\parbox{35mm}{\includegraphics[scale=0.18]{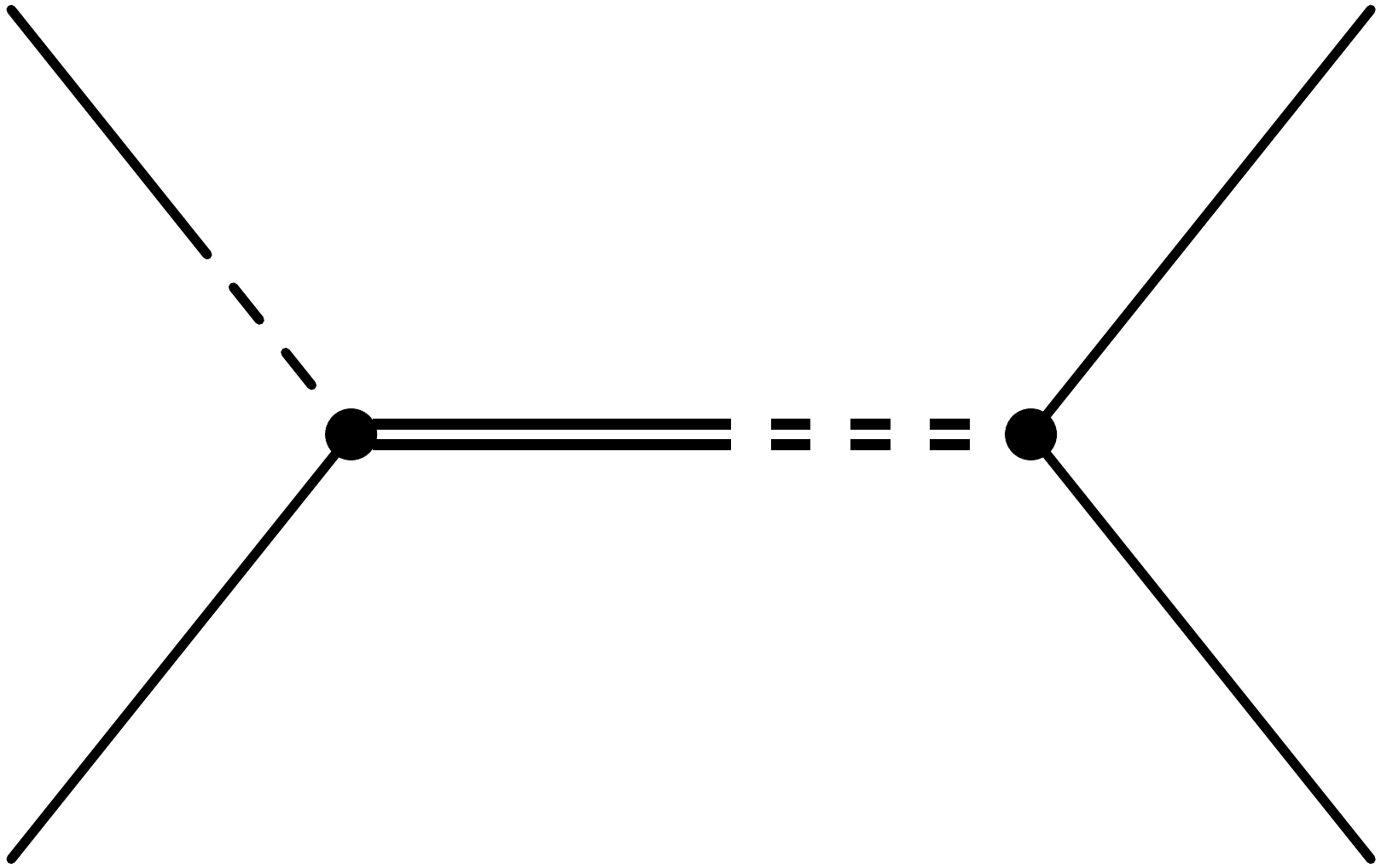} \hspace{-0.1in} \newline B}
\parbox{40mm}{\includegraphics[scale=0.18]{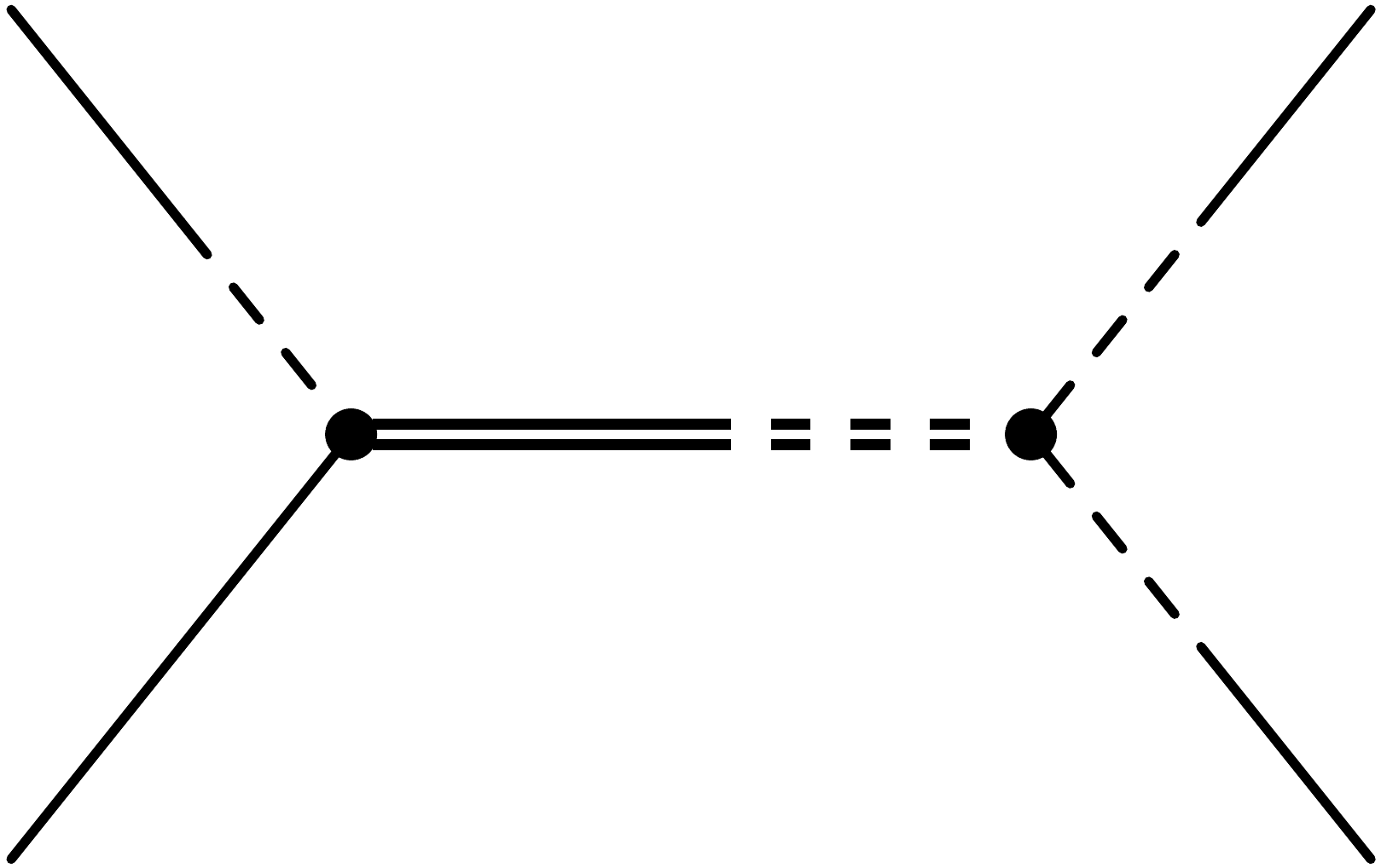} \hspace{-0.2in} \newline C} \\
\caption{Tree-level high-energy contributions to the 4-point correlation function.}
\label{tree_4pt_diags}
\end{center}
\end{figure}
\begin{figure}
\begin{center}
\hspace{-0.0in}
\parbox{35mm}{\includegraphics[scale=0.18]{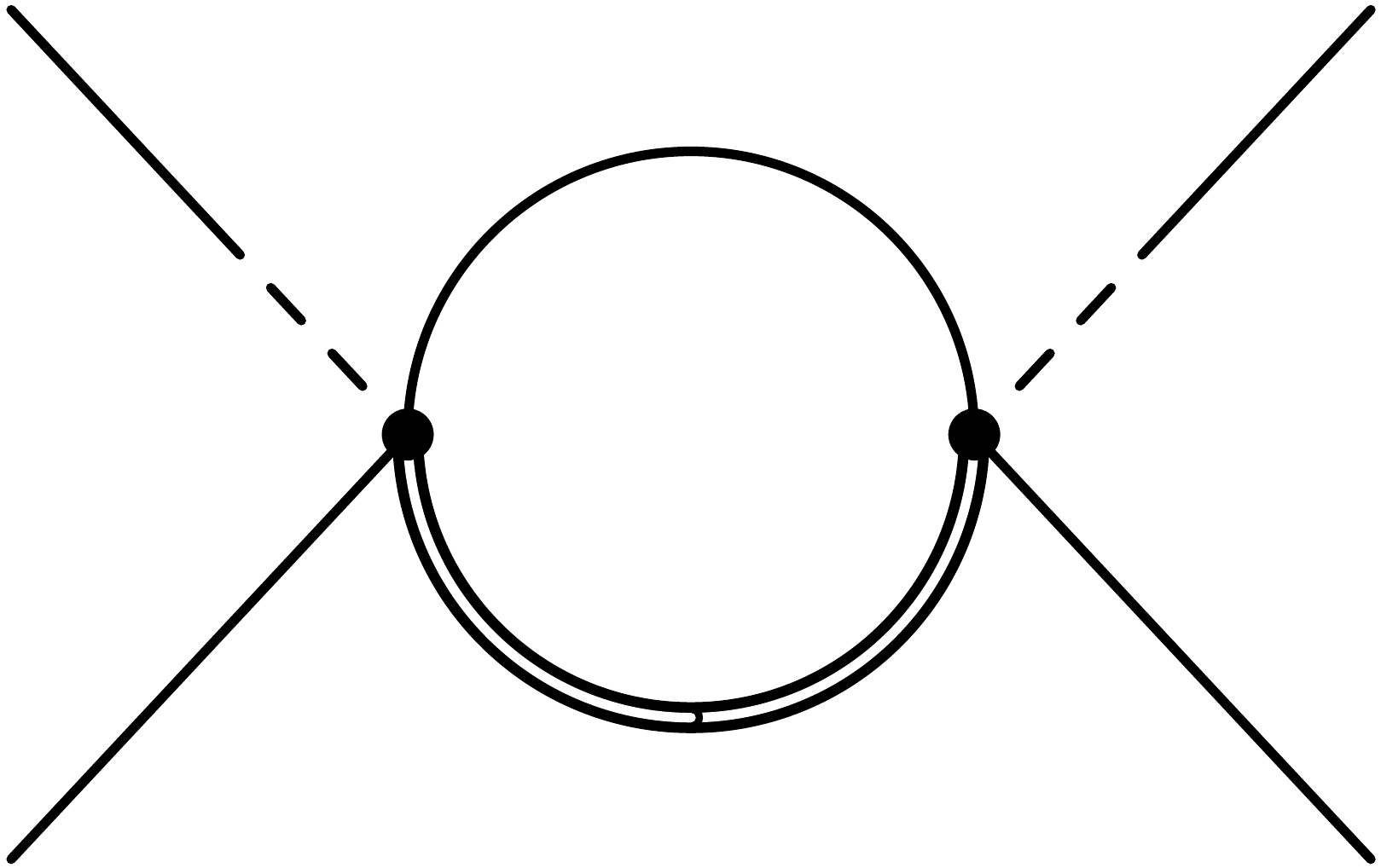} \hspace{-0.15in} \newline D}
\parbox{35mm}{\includegraphics[scale=0.18]{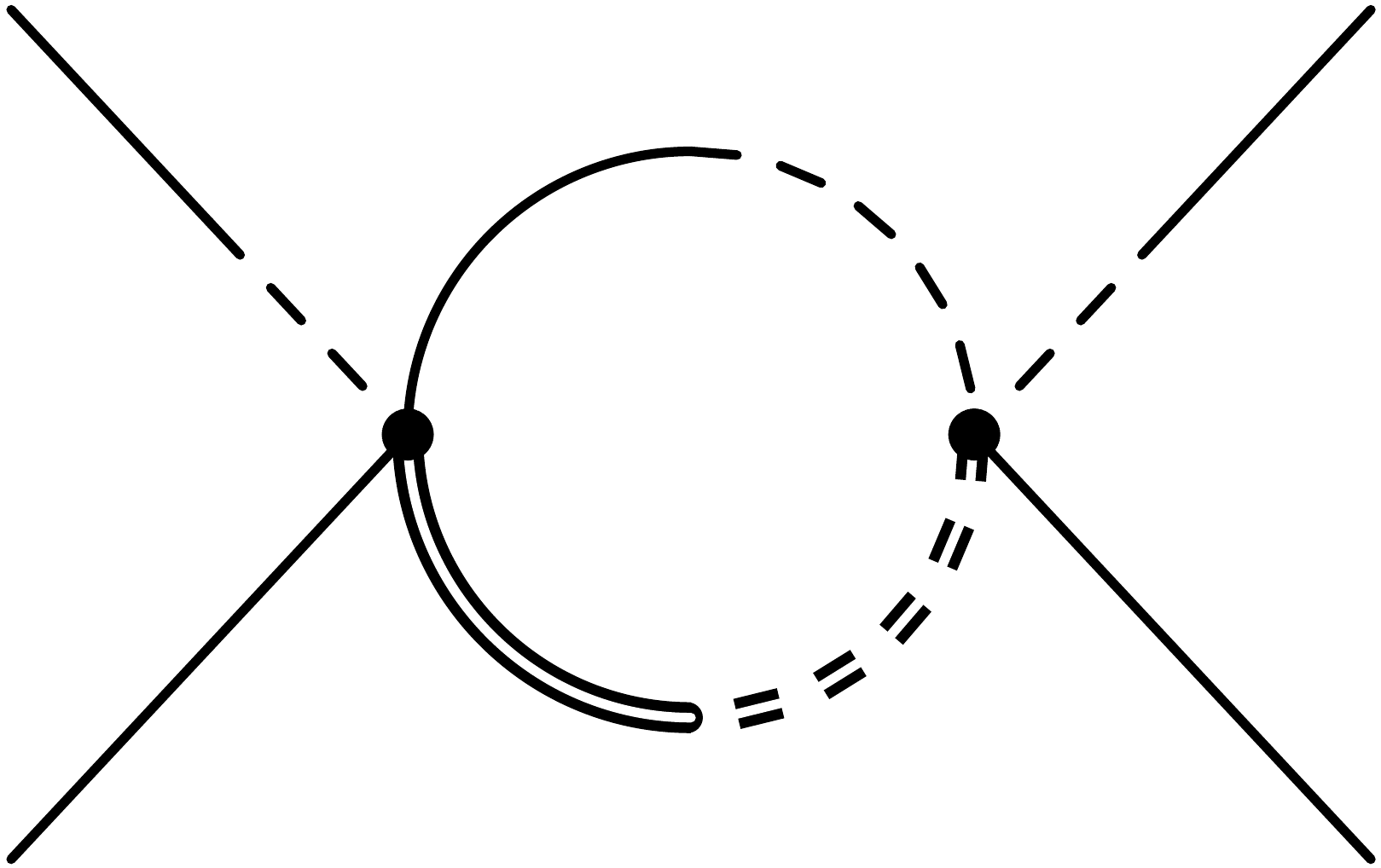} \hspace{-0.15in} \newline E} \\
\parbox{35mm}{\includegraphics[scale=0.18]{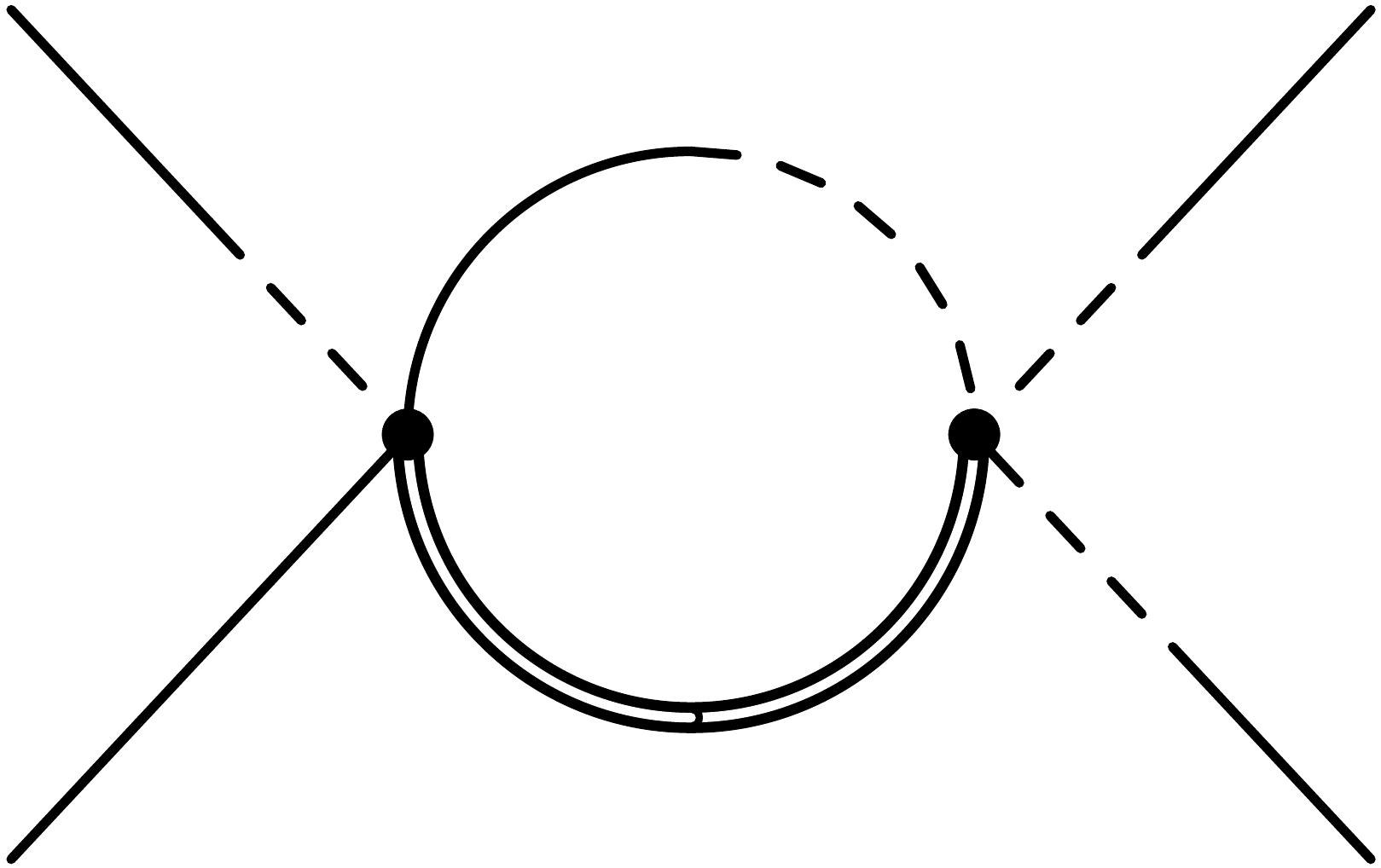} \hspace{-0.15in} \newline F}
\parbox{35mm}{\includegraphics[scale=0.18]{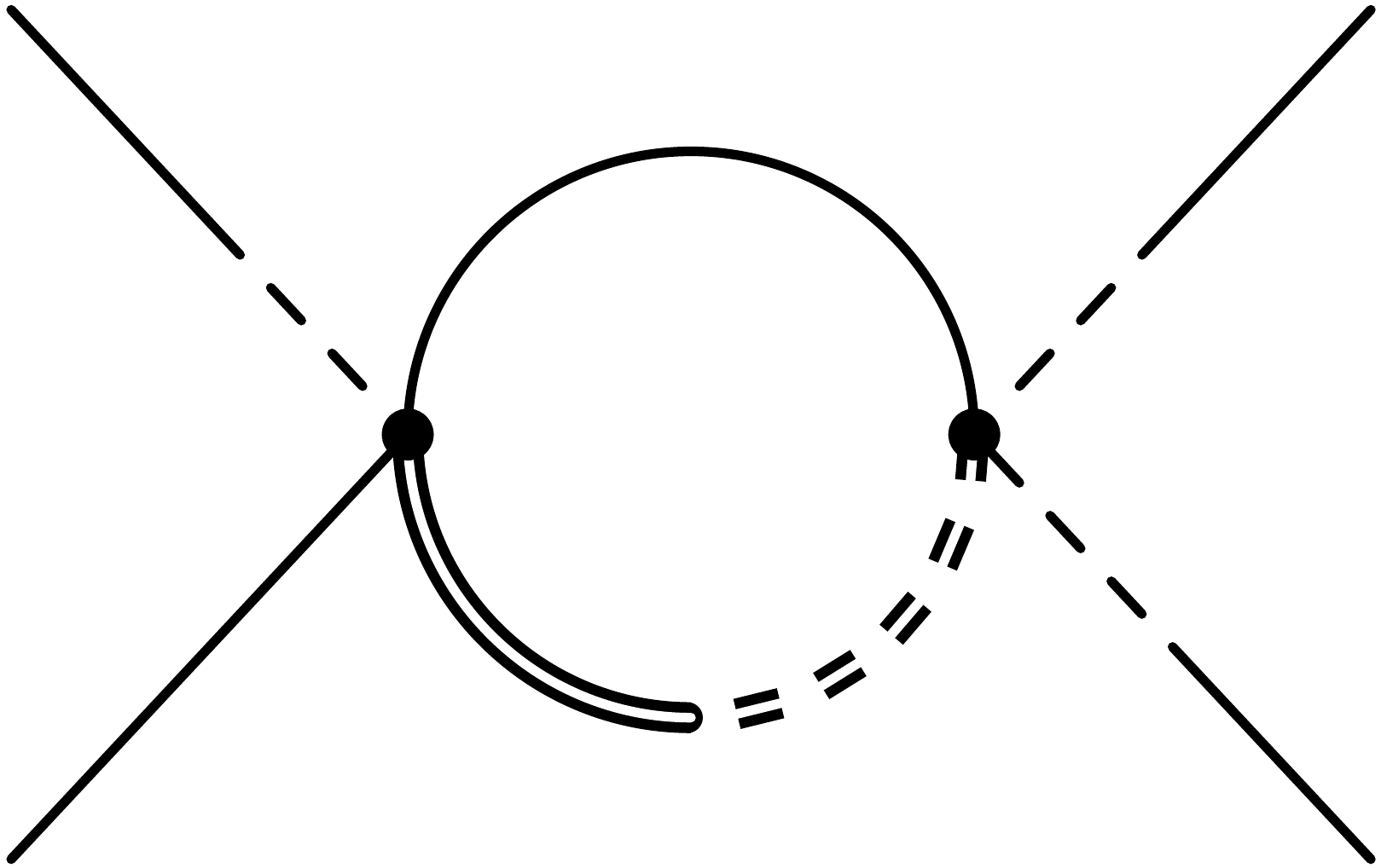} \hspace{-0.15in} \newline G} \\
\parbox{35mm}{\includegraphics[scale=0.18]{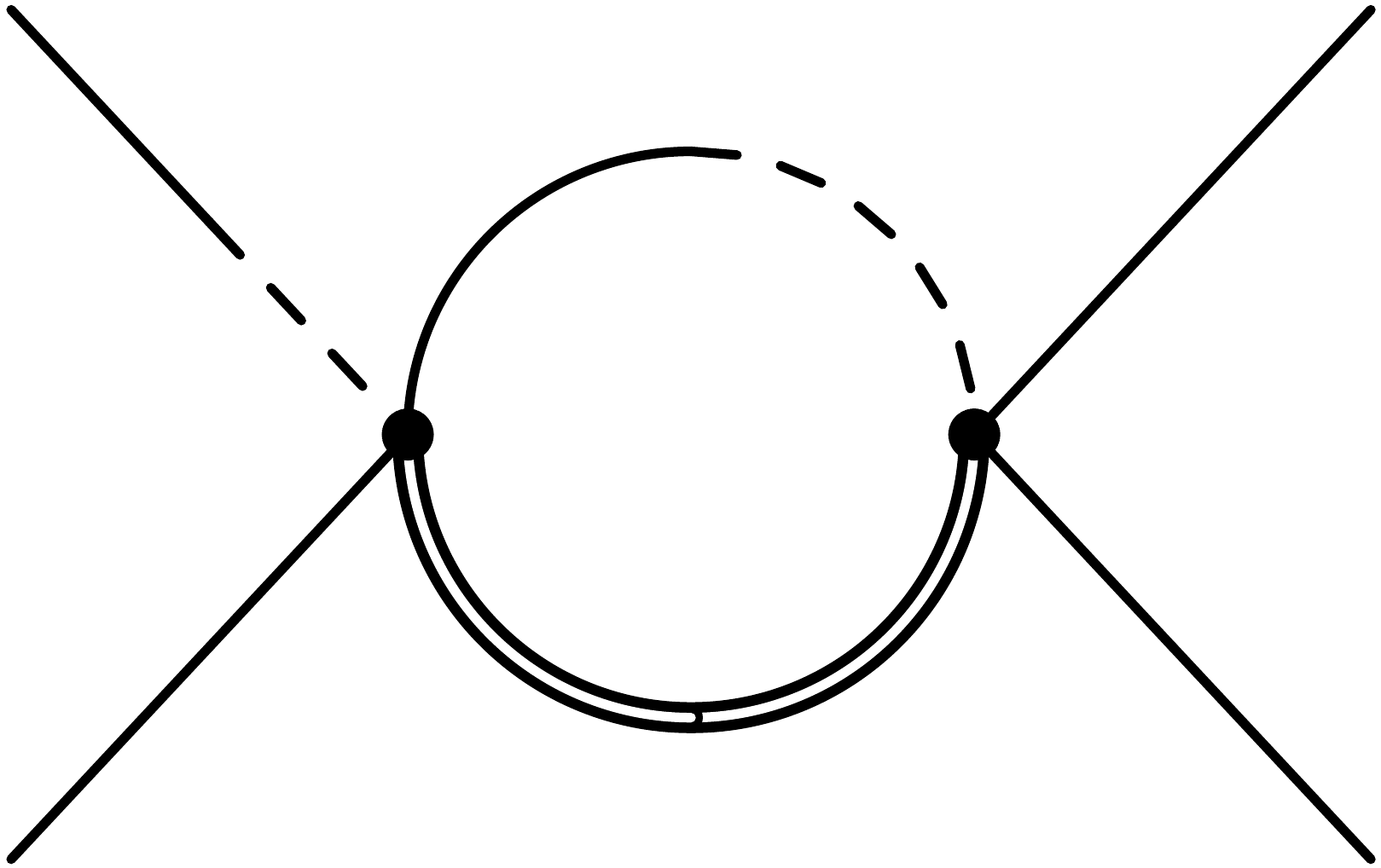} \hspace{-0.15in} \newline H}
\parbox{35mm}{\includegraphics[scale=0.18]{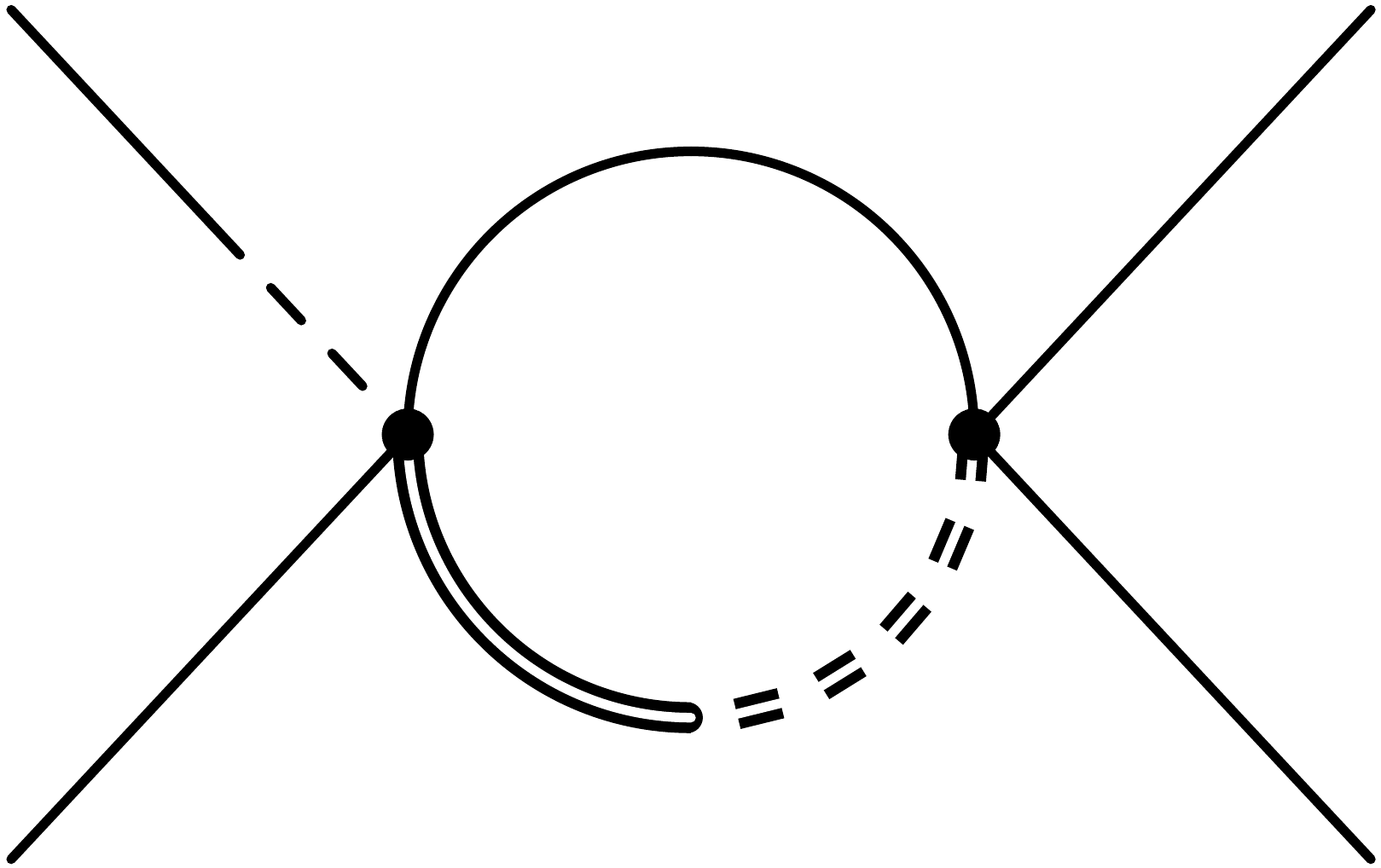} \hspace{-0.15in} \newline I}
\caption{Loop high-energy contributions to the 4-point correlation function.}
\label{loop_4pt_diags}
\end{center}
\end{figure}

Figure \ref{tree_4pt_diags} shows the three tree-level high-energy diagrams contributing to the trispectrum corrections.  We first evaluate diagram A,
\begin{eqnarray*}
T_ \varphi ^A &=& \frac{(-ig_1)^2}{(2 \pi)^3}  \int^0_{\tau_{\rm in}} d\tau_1 \ a(\tau_1)^4 \int^0_{\tau_{\rm in}} d\tau_2 \ a(\tau_2)^4 \times \\
&& \hspace{0.2in} \Big( [-iG^R_{{\bf k}_1} (0,\tau_1)] F_{{\bf k}_2}(0,\tau_1) \mathcal F_{ {\bf k}_1+{\bf k}_2}(\tau_1,\tau_2) \times \\
&& \hspace{0.3in} [-iG^A_{{\bf k}_3} (\tau_2,0)] F_{{\bf k}_4} (\tau_2,0) + \ {\rm permutations} \Big) \\
&=& \frac{g_1^2H^4}{4 (16 \pi^2 k_1 k_2 k_3 k_4)^{3/2}} \Big[ \mathcal A_1 ( {\bf k}_1, {\bf k}_2) \mathcal A_1^* ( {\bf k}_3, {\bf k}_4) + {\rm c.c.} \Big] \\
\label{osc}
&=& \frac{g_1^2 H^3}{2^{10} \pi^2(k_1 k_2 k_3 k_4)^{2} M} \times \\
&& \hspace{-0.5in} \left[ (k_1 + k_2)^2 - | {\bf k}_1 + {\bf k}_2|^2 \right]^{-1/4}  \left[ (k_3 + k_4)^2 - | {\bf k}_3 + {\bf k}_4|^2 \right]^{-1/4} \times \\
&& \hspace{-0.5in} \left( \frac{ k_1 + k_2 + \sqrt{ (k_1 + k_2)^2 - | {\bf k}_1 + {\bf k}_2|^2}}{k_3 + k_4 + \sqrt{ (k_3 + k_4)^2 - | {\bf k}_3 + {\bf k}_4|^2}} \right)^{-i M/H} + {\rm c.c.} \\
&+& {\rm permutations.}
\end{eqnarray*}
As in the bispectrum, the oscillating phase corresponds to the
difference in interaction time between the two vertices. In this case,
however, there is no loop-momentum integration to wash the
oscillations out. That these oscillations are really present is in fact easy to
understand. Due to the cosmological blueshift the difference in
interaction times translates into a difference in energies. The
Schwinger-Keldysh-Feynman diagram describes a coherent pair of heavy particles
spontaneously created each separately decaying into light particles at
different moments where the on-shell condition is reached. One can
therefore effectively think of the heavy particle as a superposition
of ``two'' states,
$|\chi\rangle=\frac{1}{\sqrt{1+|\alpha|^2}}\left(|\chi_1\rangle+\alpha|\chi_2\rangle\right)$, one state $\chi_1$ which decays to
$\chi_1 \rightarrow \varphi(k_1)+\varphi(k_2)$, the other to 
$\chi_2 \rightarrow \varphi(k_3)+\varphi(k_4)$. Moreover (though
energy is not a good quantum number in cosmology) the effective
energy encoded in the interaction time is different for $\chi_1$ and $\chi_2$.
A state which is a superposition of two different energy states with
respective energies $E_1,E_2$ and lifetimes $\tau_1,\tau_2$
famously shows oscillations in its time evolution:
\begin{eqnarray}
  \label{eq:1}
&&  \hspace{-0.4in} |\langle \chi(t)|\chi(0)\rangle|^2 = \frac{1}{(1+|\alpha|^2)^2}\left(e^{-t/\tau_1}+|\alpha|^2
    e^{-t/\tau_2} \right. \\
    \nonumber
  && \hspace{0.4in} \left. +2 |\alpha| e^{-t(\tau_1+\tau_2)/2\tau_1\tau_2}\cos t(E_2-E_1) \right).
\end{eqnarray}
This is what underlies e.g. neutrino oscillations. To emphasize that
it is the same physics here, recall that
\begin{eqnarray}
  \label{eq:2}
  \sqrt{(k_1+k_2)^2-|\mathbf{k_1}+\mathbf{k_2}|^2}=M/H|\tau_{12}|~.
\end{eqnarray}
Defining a physical energy similar to eqn (16),
\begin{eqnarray}
E_{12}=(k_1+k_2) H|\tau_{12}|,
\end{eqnarray}
$T_{\varphi}^A$ can be rewritten as
\begin{eqnarray*}
T_{\varphi}^A &\sim& \sqrt{\frac{M^2}{H^2\tau_{12}\tau_{34}}}
  \cos \left[ \frac{M}{H}\ln\frac{(E_{12}+M)\tau_{34}}{(E_{34}+M)\tau_{12}} \right] \\
&\sim& \sqrt{\frac{M^2}{H^2\tau_{12}\tau_{34}}}
  \cos \left[ \frac{M}{H}\ln\frac{\tau_{34}}{\tau_{12}} \right. \\
&& \hspace{0.4in}  \left. + \ln \left(1+\frac{E_{12}}{M} \right)-\ln \left(1+\frac{E_{34}}{M} \right) \right]
\end{eqnarray*}
Approximating $\ln(1+E/M) =E/M+\ldots$ and transforming to cosmological
time $\tau=-e^{-Ht}/H$,
\begin{eqnarray}
\nonumber
&&  T_{\varphi}^A\sim \sqrt{\frac{M^2}{H^2\tau_{12}\tau_{34}}}
  \cos \left[ M(t_{12}-t_{34})+\frac{E_{12}}{H}-\frac{E_{34}}{H} \right] \\
    \label{eq:3}
 && \hspace{0.8in} + \ \rm{permutations}
\end{eqnarray}
 we recognize the familiar $\Delta E$
interference term with a new cosmological $\Delta t$ term as well. For
the specific case $k_1+k_2=k_3+k_4$ the energy is conserved, making the vertex interaction time equal, and the oscillations disappear. This is also scale-invariant, $g^A_{\rm NL} \sim T^A_\varphi/P^3_\varphi \sim 1$.  By combining such high-energy interactions with local ones, there would be an enhancement as seen in \cite{Agullo:2011aa}.

To see whether these characteristic oscillatory features are present in the total signal, we must also compute the next two diagrams.  They are
\begin{eqnarray*} 
T_ \varphi ^B &=& \frac{(-ig_1)^2}{(2 \pi)^3} \int^0_{\tau_{\rm in}} d\tau_1 \ a(\tau_1)^4 \int^0_{\tau_{\rm in}} d\tau_2 \ a(\tau_2)^4 \times \\
&& \hspace{-0.4in} \left[ -i G^R_{{\bf k}_1} (0,\tau_1) \right] F_{{\bf k}_2}(0,\tau_1) \left[ -i \mathcal G^R_{ {\bf k}_1+{\bf k}_2}(\tau_1,\tau_2) \right] F_{{\bf k}_3} (\tau_2,0)  F_{{\bf k}_4} (\tau_2,0) \\
&& \hspace{1in} + \ {\rm permutations}, \\
T_ \varphi ^C &=& \frac{(-ig_1)^2}{(2 \pi)^3} \int^0_{\tau_{\rm in}} d\tau_1 \ a(\tau_1)^4 \int^0_{\tau_{\rm in}} d\tau_2 \ a(\tau_2)^4 \left[ -i G^R_{{\bf k}_1} (0,\tau_1) \right] \times \\
&& \hspace{-0.2in}  F_{{\bf k}_2}(0,\tau_1) \left[ -i \mathcal G^R_{ {\bf k}_1+{\bf k}_2}(\tau_1,\tau_2) \right] \left[ -i G^A_{{\bf k}_3} (\tau_2,0) \right] \left[ -i G^A_{{\bf k}_4} (\tau_2,0) \right] \\
&& \hspace{1in} + \ {\rm permutations}. 
\end{eqnarray*}
Computing their effects, one easily sees that 
\[ T_\varphi^B = T_\varphi^C = - \frac{1}{2} T_\varphi^A. \]
The leading $H/M$ contribution therefore vanishes. 
\subsection{Loop High-Energy Corrections}
Now turning to the loop corrections in Figure \ref{loop_4pt_diags}, the first is
\begin{eqnarray*}
T_ \varphi ^D &=& \frac{(-i\lambda_1)^2}{(2 \pi)^3} \int^0_{\tau_{\rm in}} d\tau_1 \ a(\tau_1)^4 \int^0_{\tau_{\rm in}} d\tau_2 \ a(\tau_2)^4 \times \\
&& \hspace{-0.4in} \int \frac{ d^3 {\bf q}}{( 2\pi)^3} \Big( [-iG^R_{{\bf k}_1} (0,\tau_1) ] F_{{\bf k}_2}(0,\tau_1) F_{{\bf q}}(\tau_1,\tau_2) \times \\
&& \hspace{-0.5in} \mathcal F_{{\bf k}_1+{\bf k}_2+{\bf q}} (\tau_1,\tau_2) [-iG^A_{{\bf k}_3} (\tau_2,0) ] F_{{\bf k}_4}(\tau_2,0) + {\rm perms.} \Big) \\
&& \hspace{-0.5in} = \frac{\lambda_1^2 H^4}{2 \pi^3 (k_1 k_2 k_3 k_4)^{3/2}} \times \\
&& \hspace{-0.4in} \int \frac{d^3 {\bf q}}{(2 \pi)^3} \Big[ \mathcal B_1 ( {\bf k}_1, {\bf k}_2, {\bf q}) \mathcal B_1^* ( {\bf k}_3, {\bf k}_4, {\bf q}) + {\rm c.c.} + {\rm perms.} \Big] .
\end{eqnarray*}
There is a rapidly-varying phase present completely analogous to (\ref{phase}),
\[ \left( \frac{ k_1 + k_2 + q + \sqrt{ (k_1 + k_2+q)^2 - | {\bf k}_1 + {\bf k}_2 + {\bf q}|^2}}{k_3 + k_4 + q + \sqrt{ (k_3 + k_4 + q)^2 - | {\bf k}_3 + {\bf k}_4 + {\bf q}|^2}} \right)^{-i M/H} . \]
As in the bispectrum loop integration, the integration over ${\bf q}$ will only allow contributions near $k_1 + k_2 = k_3 + k_4$.  Now defining variables appropriate to this case,
\begin{eqnarray*}
 \kappa_{12} &\equiv& k_1 + k_2 - (k_3 + k_4), \\
u^{-1} &\equiv& - \sqrt{  (k_1+k_2+q)^2 - | {\bf k}_1 + {\bf k}_2 + {\bf q}|^2 }, \\
E &\equiv& M q |u| .
\end{eqnarray*}
Taking $\theta$ to be the angle between $k_{12} \equiv {\bf k}_1 + {\bf k}_2$ and ${\bf q}$, the $(u,E)$ variables can be inverted to yield
\begin{eqnarray*}
q &=& \frac{E }{M|u|}, \\
1- \cos \theta &=& \frac{M |u|}{2k_{12} E } \left[ |u|^{-2}  - (k_1 + k_2)^2 + k_{12}^2 \right].
\end{eqnarray*}
Just as in the bispectrum case, the integral transforms as
\begin{equation}  
\hspace{-0.0in} \int \frac{d ^3 \bf q}{(2 \pi)^3} \rightarrow - \frac{1}{(2 \pi)^2 M^2 k_{12}} \int \frac{du E d E}{u^5 } . 
\end{equation}
The integrand transforms as
\[ \mathcal B_1 ( {\bf k}_1, {\bf k}_2, {\bf q}) \mathcal B_1^* ( {\bf k}_3, {\bf k}_4, {\bf q})  \rightarrow  \frac{\pi |u|^4 M^2 e^{- i \frac{M}{H} \kappa_{12} u } }{16 H E \sqrt{k_1 k_2 k_3 k_4} } + \mathcal O(\kappa_{12}) . \]
The bounds are now
\begin{eqnarray*}
&& \hspace{-0.4in} \frac{M |u| \left[ |u|^{-2} + k_{12}^2 - (k_1 + k_2)^2 \right]}{2 (k_1 + k_2 +  k_{12})} \leq E \leq \Lambda - M(k_1 + k_2)|u| , \\
 && \hspace{-0.2in} |u_{\pm}|  = \frac{1}{ \left( k_{12}+ k_1 + k_2 \right)^2} \times \\
 && \hspace{0.0in} \left[ \left( \frac{\Lambda}{M} \right) k_{12} \pm \sqrt{ \left( \frac{\Lambda}{M} \right)^2  k_{12}^2 -  \left(k_{12} + k_1 + k_2 \right)^2} \right] .
\end{eqnarray*}
The energy integral is again trivial, and expanding near the extremum at $|u_0| = M/\Lambda (k_1 + k_2 + k_{12})$ the resultant integrand can be approximated as Gaussian:
\begin{eqnarray*}
&& \hspace{-0.2in} |u|^{-1} \Big[ \Lambda - M(k_1 + k_2)|u| \\
&& \left. \hspace{0.5in} - \frac{M |u| \left[ |u|^{-2} + k_{12}^2 - (k_1 + k_2)^2 \right]}{2 (k_1 + k_2 +  k_{12})} \right] \\
&\approx& \frac{ \Lambda^2 (k_1 + k_2 + k_{12})}{2M} \times \\
&& \exp \left[ - \frac{1}{2} \left( \frac{\Lambda }{M} \right)^{2} (k_1 + k_2 +  k_{12})^2 (u-u_0)^2\right].
\end{eqnarray*}
Now performing the integral over $u$,
\begin{eqnarray*}
&& \hspace{-0.1in} \int_{- \infty}^{\infty} du \ e^{ -i \frac{M}{H} \kappa_{12} u  - \frac{1}{2} \left( \frac{\Lambda }{M} \right)^{2} (k_1 + k_2 +  k_{12})^2 (u-u_0)^2 } + {\rm c.c.} \\
&\approx& \frac{  \sqrt{  2 \pi} M}{\Lambda( k_1 + k_2 +  k_{12}) }  e^{ - i \frac{M}{H} \kappa_{12} u_0 - \frac{1}{2} \left( \frac{M^2  \kappa_{12}}{H \Lambda (k_1 + k_2 +  k_{12})} \right)^2 } + {\rm c.c.} \\
&\approx& \frac{4 \pi H}{M} \delta(\kappa_{12}) .
\end{eqnarray*}
The final answer is then
\begin{equation}
\label{ta}
T^D_{\varphi} = \frac{\pi \lambda_1^2 H^4 \Lambda^2 (k_1 + k_2 +  k_{12})}{(2 \pi)^4 M^2 (k_1 k_2 k_3 k_4)^2 k_{12}} \delta( k_1 + k_2 - k_3 - k_4) + {\rm perms.}
\end{equation} 
The $(\Lambda/M)^2$ can be absorbed into the bare coupling $\lambda_1$.  Unlike the tree-level trispectrum corrections, there is no $H/M$ dependence in the magnitude of (\ref{ta}). Like the other high-energy diagrams calculated, it is scale-invariant.
 
The next diagram is similar,
\begin{eqnarray*}
T_ \varphi ^E &=& \frac{(-i\lambda_1)^2 }{(2 \pi)^3} \int^0_{\tau_{\rm in}} d\tau_1 \ a(\tau_1)^4 \int^0_{\tau_{\rm in}} d\tau_2 \ a(\tau_2)^4 \times \\
&& \hspace{-0.4in} \int \frac{ d^3 {\bf q}}{( 2\pi)^3}  \Big( [-i G^R_{{\bf k}_1} (0,\tau_1) ] F_{{\bf k}_2}(0,\tau_1) [-iG^R_{{\bf q}}(\tau_1,\tau_2)] \times \\
&& \hspace{-0.5in} \mathcal  [-iG^R_{{\bf k}_1+{\bf k}_2+{\bf q}} (\tau_1,\tau_2) ][-i G^A_{{\bf k}_3} (\tau_2,0) ] F_{{\bf k}_4}(\tau_2,0) + {\rm perms.} \Big).
\end{eqnarray*}
It is easy to see that $T^E_\varphi = - \frac{1}{2} T^D_\varphi$.  Using the formula for $u_c$ in the $\mathcal B$'s, we assume that $k_1 + k_2 \geq k_3 + k_4$ to ensure $u_1 \geq u_2$.  

Finally, $T_ \varphi ^F$ and $T_ \varphi ^G$ cancel, and $T_ \varphi ^H$ and $T_ \varphi ^I$ cancel.  
\subsection{Total Trispectrum}
Taking advantage of the cancellations between several diagrams, the total trispectrum is
\begin{eqnarray*}
T_\varphi &=& (2 \pi)^6 \delta^3( {\bf k}_1 - {\bf k}_2) \delta^3( {\bf k}_3 - {\bf k}_4)P_\varphi(k_1) P_\varphi(k_3) \\
&& \hspace{-0.3in} + {\rm permutations}  +  T^{\rm self}_\varphi + T_\varphi^D +  T_\varphi^E.
\end{eqnarray*}
While oscillations arise at order $H/M$ in individual diagrams, their cancellation means that no such feature appears in the final result.  

\section{Conclusion}
We have computed the \emph{generic} primordial bi- and trispectrum corrections in an inflating background due to high energy physics.  The dominant physics arises from the New Physics Hypersurface where the cosmological blueshift can just create an on-shell heavy particle.  As predicted \cite{Holman:2007na, Meerburg:2009ys, Meerburg:2009fi}, the bispectrum corrections peak near the elongated shape of momentum-triangle; here we derive this \emph{ab initio}.  The bispectrum otherwise contains no extraordinary features.  The trispectrum has no leading order effect.  Even though the interference effect of the effective mass-shell-time-difference for a pair of heavy particles, similar to neutrino oscillations, this cancels when all diagrams are taken into account.  

Like that of the power spectrum, the magnitude of the bispectrum high-energy corrections are estimated to be of order $H/M$.  For optimistic but not impossible values of $H/M \sim 0.01$ these corrections could possibly be measured by \emph{Planck} \cite{planck} and other precision experiments in the near future.  Also like the power spectrum, the leading corrections arising from the dynamical part of the effective action cancel and the dominant corrections arise entire from the density matrix.  

We emphasize that we have only calculated the model-independent corrections.  Specific models will no doubt produce very rich features and should be studied in more detail. 
\section{Acknowledgments}
We would like to thank F.~Bouchet, J.~Fergusson, E.~Komatsu, and B.~Wandelt for discussions.  This research was supported in part by a VIDI and a VICI Innovative Research Incentive Award from the Netherlands Organisation for Scientific Research (NWO), a van Gogh grant  from the NWO, and the Dutch Foundation for Fundamental Research on Matter (FOM).

\end{document}